\batchmode
\makeatletter
\def\input@path{{\string"C:/Users/USER/Google Drive/draft_of_Numerical_approach_to_resonant_transmission_and_relaxation/inbaloz_paper_folder/\string"}}
\makeatother
\documentclass[12pt,english]{article}
\usepackage{helvet}
\usepackage[T1]{fontenc}
\usepackage[latin9]{inputenc}
\usepackage{geometry}
\usepackage{float}
\usepackage{amsmath}
\usepackage{amssymb}
\usepackage{mathdots}
\usepackage{graphicx}
\usepackage{setspace}
\usepackage{esint}
\makeatletter
\numberwithin{equation}{section}
\newcommand{\lyxaddress}[1]{
	\par {\raggedright #1
	\vspace{1.4em}
	\noindent\par}
}

\usepackage{braket}

\usepackage{pdfcomment}

\makeatother

\usepackage{babel}
\begin{document}
\date{}

\title{Evaluation of Dynamical Properties of Open Quantum Systems Using
the Driven Liouville-von Neumann Approach: Methodological Considerations}

\author{Inbal Oz,\textsuperscript{1,2} Oded Hod,\textsuperscript{1,2} and
Abraham Nitzan\textsuperscript{1,2,3}}
\maketitle

\lyxaddress{\textsuperscript{1}Department of Physical Chemistry, School of Chemistry,
The Raymond and Beverly Sackler Faculty of Exact Sciences, Tel Aviv
University, Tel Aviv, IL 6997801}

\lyxaddress{\textsuperscript{2}The Sackler Center for Computational Molecular
and Materials Science, Tel Aviv University, Tel Aviv, IL 6997801}

\lyxaddress{\textsuperscript{3}Department of Chemistry, University of Pennsylvania,
Philadelphia, PA, USA 19103}
\begin{abstract}
Methodological aspects of using the driven Liouville-von Neumann (DLvN)
approach for simulating dynamical properties of molecular junctions
are discussed. As a model system we consider a non-interacting resonant
level uniformly coupled to a single Fermionic bath. We demonstrate
how a finite system can mimic the depopulation dynamics of the dot
into an infinite band bath of continuous and uniform density of states.
We further show how the effects of spurious energy resolved currents,
appearing due to the approximate nature of the equilibrium state obtained
in DLvN calculations, can be avoided. Several ways to approach the
wide band limit that is often adopted in analytical treatments, using
a finite numerical model system are discussed including brute-force
increase of the lead model bandwidth as well as efficient cancellation
or direct subtraction of finite-bandwidth effect. These methodological
considerations may be relevant also for other numerical schemes that
aim to study non-equilibrium thermodynamics via simulations of open
quantum systems.
\end{abstract}

\section{Introduction}

The study of electron dynamics and conductance in small electronic
systems coupled to one or more free-electron reservoirs (each in its
own equilibrium but not necessarily at equilibrium with each other)
has attracted much attention over the past decade due to its importance
for studies in the fields of molecular electronics \cite{nitzan_electron_2003,cuevas_molecular_2010,ghosh_nanoelectronics_2016}
spectroscopy \cite{galperin_molecular_2012}, and quantum thermodynamics
\cite{bruch_quantum_2016,ludovico_dynamical_2014,ludovico_dynamics_2016,esposito_quantum_2015,tuovinen_phononic_2016,gelbwaser-klimovsky_thermodynamics_2015,arrachea_microscopic_2012,kosloff_quantum_2014}.
Although substantial advances in observing and describing such processes
were made in the past three decades, the study of particle and energy
transfer in processes dominated by resonance transmission between
non-equilibrium environments remains a major experimental and theoretical
challenge. As in other dynamical problems, computer simulations may
offer a complementary approach to purely theoretical analysis in this
field.

A major challenge for modeling electronic transport through such nanometric
structures is the ability to provide an appropriate non-equilibrium
description of the entire (infinite in principle) system. This problem
is often solved by replacing the full system by a finite system with
proper account of the non-equilibrium open boundaries. One of the
most widely used approaches to address this challenge is the extension
of the Landauer formalism to address dynamical effects by using the
non-equilibrium Green's function method \cite{haug_quantum_2008}.
This method can provide analytical solutions for non-interacting models
\cite{chen_simple_2014,cheng_simulating_2006,di_ventra_transport_2004}
but it becomes computationally demanding for steady-states of more
realistic model systems as well as for systems affected by time-dependent
driving.

An alternative approach to describing electronic transport in such
systems is the use of numerical simulations. Since simulated models
are necessarily of finite size, ways to impose the infinite (in principle)
nature of the real system need to be devised. In vacuum scattering
problems this is usually achieved by using absorbing boundary conditions
\cite{seideman_calculation_1992}. When the environment of the simulated
system consists of metallic leads with occupied electronic states,
the numerical boundary has to account for both electron absorption
and injection. A variety of methods, too extensive to detail herein,
have been developed for this purpose considering both model Hamiltonians
\cite{baer_quantum_1997,koch_femtosecond_2003,galperin_current-induced_2005,kleinekathofer_switching_2006,fainberg_light-induced_2007,katz_stochastic_2008,subotnik_nonequilibrium_2009,rothman_nonequilibrium_2010,volkovich_transient_2011,renaud_time-dependent_2011,peskin_coherently_2012,nguyen_how_2015}
and realistic model systems \cite{baer_ab_2003,baer_ab_2004,di_ventra_transport_2004,bushong_approach_2005,sanchez_molecular_2006,cheng_simulating_2006,zheng_time-dependent_2007,evans_dynamic_2009,ercan_tight-binding_2010,zheng_time-dependent_2010,xing_first-principles_2010,ke_time-dependent_2010,wang_time-dependent_2013,schaffhauser_using_2016}.
Among the latter, the recently proposed Driven Liouville-von Neumann
(DLvN) approach \cite{zelovich_state_2014,zelovich_molecule-lead_2015,zelovich_driven_2016,zelovich_parameter-free_2017},
imposes the required boundary conditions by augmenting the Liouville-von
Neumann (LvN) equation of motion with non-unitary source and sink
terms. The latter drive each lead towards an equilibrium state determined
by the chemical potential and electronic temperature of the implicit
bath to which it is coupled \cite{zelovich_state_2014,zelovich_molecule-lead_2015,zelovich_driven_2016,zelovich_parameter-free_2017}.
When the driving enforces different equilibrium states on different
leads, the DLvN method was shown to provide a reliable representation
of the electronic transport problem, closely reproducing the Landauer
formalism results at steady-state \cite{zelovich_state_2014,zelovich_driven_2016}.
Furthermore, it was shown that, for non-interacting systems \cite{hod_driven_2016,gruss_landauers_2016,elenewski_communication:_2017},
the DLvN equation of motion can be recast into Lindblad form, thus
it inherently preserves density matrix positivity \cite{subotnik_nonequilibrium_2009,zelovich_driven_2016,elenewski_communication:_2017}.
Notably, within this approach, external dynamic perturbations such
as alternating bias voltages, varying gate potentials, and time-dependent
external fields may be readily imposed. These can drive the system
out of its equilibrium or steady states \cite{chen_simple_2014} and
invoke intriguing physical phenomena that are manifested in the dynamical
properties of the system, beyond the scope of the well established
equilibrium thermodynamic theory.

A simplistic model that can demonstrate such effects is a resonant
level model, where a single non-interacting spinless state (often
referred to as a quantum dot) is coupled to a manifold of non-interacting
spinless lead states. Here, shifting the position of the dot level
with respect to the chemical potential of the lead states mimics the
application of an external time-dependent gate potential on the dot.
Recent analytical analysis of this model, at the wide band limit (WBL)
\cite{verzijl_applicability_2013,baldea_invariance_2016,covito_transient_2018},
enabled the calculation of thermodynamic functions to first order
beyond the quasistatic (QS) limit. These were shown to fulfill the
first and second laws of thermodynamics and reproduce the equilibrium
and weak coupling results in the appropriate limits \cite{bruch_quantum_2016}.
Such treatments, however, often rely on perturbation theory and hence
are limited to cases where a small parameter can be identified. Specifically,
in the treatment mentioned above, the dot level driving rate was taken
to be considerably smaller than the typical internal relaxation rate
of the lead. Hence, to gain access to dynamical regions that are beyond
the reach of current analytical treatments of this (and more complex)
model one can harness the numerical flexibility of the DLvN approach.
Nevertheless, care should be taken with the practical implementation
of the numerical simulation so as to ensure that the results, necessarily
obtained for a finite lead model, faithfully represent the desired
physical properties of an infinite environment at equilibrium. As
general guidance, some rules of thumb have been introduced to assess
the finite lead model size required to verify the validity of the
Markovian approximation adopted in the DLvN approach \cite{subotnik_nonequilibrium_2009,elenewski_communication:_2017}.
When numerical convergence with respect to the finite lead model size
is achieved its discrete spectrum mimics well the continuous density
of state of the corresponding (semi-)infinite bath. However, when
using numerical simulations to extend analytical models towards new
dynamical regimes, one should also keep in mind that simplifying assumptions,
such as the wide band approximation (WBA), that are often invoked
in approximate analytical treatments, are not always readily transferable
to the numerical calculation. The present manuscript addresses these
and related methodological aspects of using numerical simulations
in general and, in particular, the DLvN approach to complement analytical
treatments in the study of particle and heat fluxes through molecular
interfaces.

\section{Relaxation Dynamics\label{Sec: relaxation dynamics}}

To set the stage for demonstrating important methodological aspects
of using numerical schemes to simulate non-equilibrium scenarios,
we consider first the simple relaxation dynamics of an initially occupied
dot level coupled to an empty manifold of lead levels. The non-interacting
Hamiltonian of the entire system is given by

\begin{equation}
\hat{H}(t)=\hat{H}_{d}\left(t\right)+\hat{H}_{L}+\hat{H}_{V},\label{eq: H of driven resonant level}
\end{equation}

where $\hat{H}_{d}(t)=\varepsilon_{d}\left(t\right)c_{d}^{\dagger}c_{d}$
is the Hamiltonain of the dot, $\hat{H}_{L}=\sum_{l}\varepsilon_{l}c_{l}^{\dagger}c_{l}$
represents the lead section, and $\hat{H}_{V}=\hat{V}_{dL}+\hat{V}_{Ld}$
is the dot/lead coupling term, where $\hat{V}_{dL}=\sum_{l}\left(V_{l}d^{\dagger}c_{l}\right)$
and $\hat{V}_{Ld}=\hat{V}_{dL}^{\dagger}$. Here, $\varepsilon_{d}\left(t\right)$
and $\varepsilon_{l}$ are the energies of the dot and lead level
$l$, respectively, and $c_{i}^{\dagger}$ and $c_{i}$ are the creation
and annihilation operators for an electron in level $i=d,l$. Note
that, for the sake of simplicity, we have assumed that the only time-dependence
in the Hamiltonian stems from shifts in the position of the dot level,
with no effect on the lead levels and the lead-dot coupling terms.
Naturally, within the numerical treatment discussed below, this simplifying
assumption can be readily lifted.

\subsection{Analytical Solution}

Under the WBA, where the manifold of lead states is assumed to be
of infinite width with a uniform and continuous density of states,
$\rho$, this model has a fully analytical solution. Upon uniformly
coupling the isolated dot level to the manifold of lead states, its
original delta-function shape, $\delta(\varepsilon-\varepsilon_{d})$,
broadens into a Lorentzian function of the form: 

\begin{equation}
A(\varepsilon-\varepsilon_{d};\Delta,\gamma)=\frac{1}{\pi}\frac{\hbar\gamma/2}{(\varepsilon-\varepsilon_{d}-\Delta)^{2}+(\hbar\gamma/2)^{2}}.\label{eq: A lorantzian}
\end{equation}

Here, $\hbar=h/\left(2\pi\right)$ is the reduced Plank's constant,
$\Delta$ represents the shift in the dot's level position due to
the coupling to the lead, and $\hbar\gamma$ is the level broadening
due to its finite lifetime. The latter corresponds the leakage rate
of particles from the dot to the lead manifold as given by Fermi's
golden rule \cite{nitzan_chemical_2006}:

\begin{equation}
\gamma\left(\varepsilon\right)=\frac{2\pi}{\hbar}\sum_{l}\left|V_{l}\right|^{2}\delta\left(\varepsilon-\varepsilon_{l}\right),\label{eq: gm Fermi}
\end{equation}

which at the WBL, assuming an infinite lead band of constant density
of states, $\rho$, and uniform coupling to the dot ,$V_{l}=V=const$,
can be evaluated as:

\begin{equation}
\gamma^{WBA}=\frac{2\pi}{\hbar}\intop_{-\infty}^{\infty}\left|V_{l}\right|^{2}\rho\left(\varepsilon_{l}\right)\delta\left(\varepsilon-\varepsilon_{l}\right)d\varepsilon_{l}=\frac{2\pi}{\hbar}\left|V\right|^{2}\rho\intop_{-\infty}^{\infty}\delta\left(\varepsilon-\varepsilon_{l}\right)d\varepsilon_{l}=\frac{2\pi}{\hbar}\left|V\right|^{2}\rho.\label{eq: gm Fermi WBA}
\end{equation}

\begin{figure}
\begin{centering}
\includegraphics[scale=0.75]{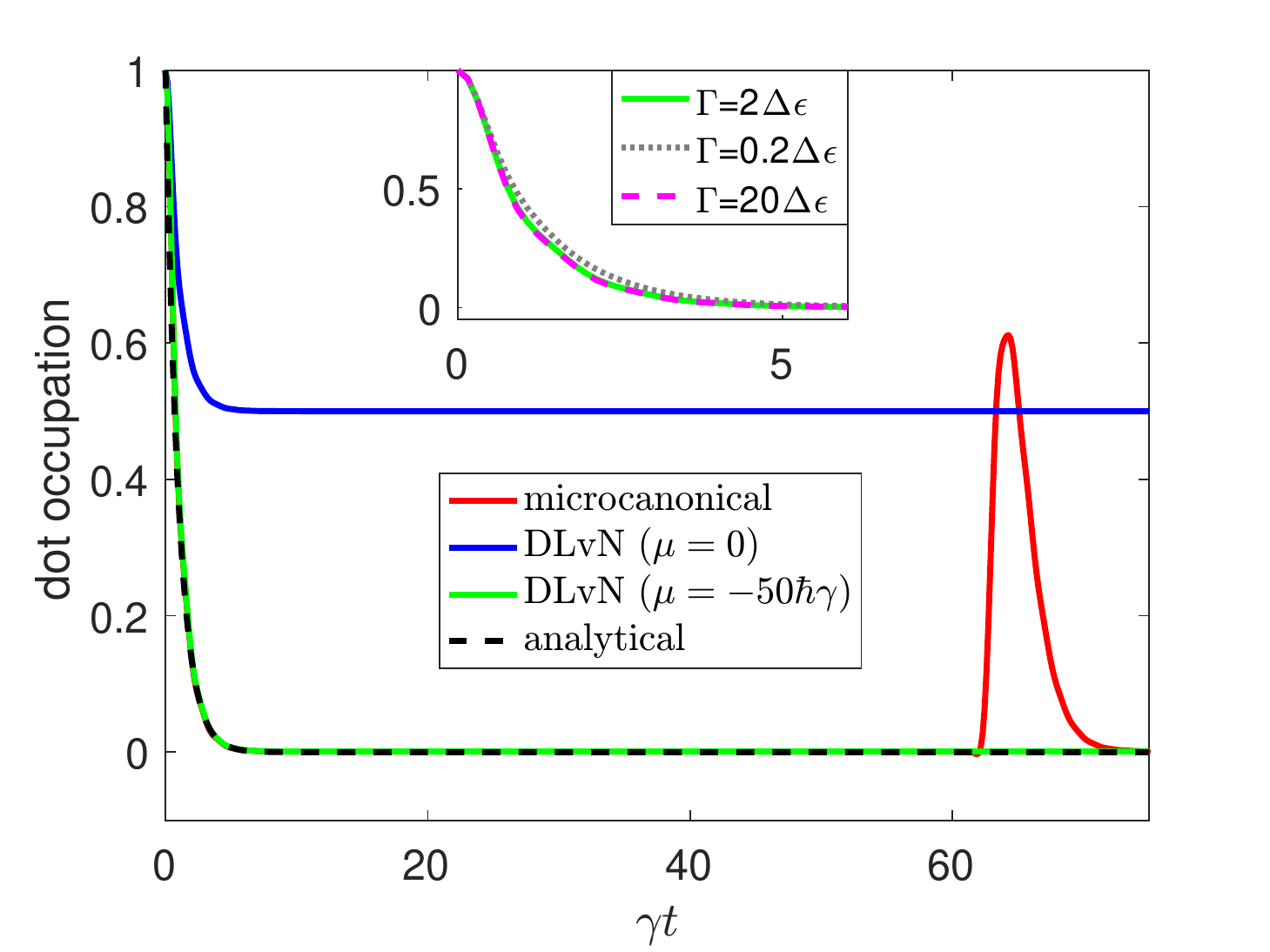}
\par\end{centering}
\caption{Dot depopulation dynamics as calculated using (i) the analytical WBA
treatment (dashed black line); (ii)\textbf{ }microcanonical simulations
(full red line); (iii) DLvN simulations with an empty (green) and
half filled (blue) lead. System parameters are provided in the main
text. \label{fig: pointcare} Inset: Driving rate sensitivity test
of the DLvN dynamics calculated with $\Gamma=0.2\Delta\varepsilon/\hbar$
(dashed gray), $\Gamma=2\Delta\varepsilon/\hbar$ (green), and $\Gamma=20\Delta\varepsilon/\hbar$
(dashed magenta). }
\end{figure}

By virtue of the Fourier transform of its Lorentzian spectral function,
the population of an initially occupied dot level $(\psi\left(t=0\right)=d^{\dagger}\left|0\right\rangle )$
decays exponentially with time into the empty manifold of lead levels
($P_{d}\left(t\right)=\left|\left\langle \psi\left(t=0\right)\right|e^{-\frac{i}{\hbar}\hat{H}t}\left|\psi\left(t=0\right)\right\rangle \right|^{2}=e^{-\gamma t}$)
with a characteristic decay time of $1/\gamma$ as indicated by the
dashed black line in Fig. \ref{fig: pointcare}.

\subsection{Closed System Numerical Treatment}

The most straightforward numerical approach to simulate this temporal
behavior of the dot's population is the microcanonical scheme \cite{di_ventra_transport_2004}.
Here, the infinite (in principle) system is represented by a finite
model consisting of the dot level uniformly coupled to a finite set
of lead levels. Two important differences between this model and the
one used for the analytical treatment above should be noted: (i) the
density of lead states is discrete, and (ii) the band of lead states
is of finite width. Nevertheless, we expect that when the lead manifold
is sufficiently dense and the position of the dot is far enough from
the band edges, the numerical simulation will reproduce the short
time dynamics of the analytical treatment. To demonstrate this, we
choose a finite lead model consisting of $N_{L}=100$ equispaced levels
that span a bandwidth of $W=10\hbar\gamma$. The corresponding level
spacing and density of states are thus given by $\Delta\varepsilon=\rho^{-1}=\frac{W}{N}=0.1\hbar\gamma$,
respectively. The dot energy, $\varepsilon_{d}$, is positioned at
the center of the lead's band and is uniformly coupled to all lead
levels via a coupling constant of $V=\sqrt{\frac{\hbar}{2\pi}\frac{\gamma}{\rho}}=\frac{\hbar\gamma}{\sqrt{20\pi}}$
(see Eq. \ref{eq: gm Fermi WBA}). We note that, the choice of the
specific value of $\gamma$ (chosen herein such that $\hbar\gamma=0.1$
eV) is arbitrary as the results presented below are scalable with
respect to it. Hence, we set all other parameters in terms of $\gamma$
and present the result in unitless format. The dynamics of the system
is simulated via the LvN equation of motion for the single-particle
density matrix of the system, $\hat{\sigma}(t)$:
\begin{equation}
\frac{d}{dt}\hat{\sigma}\left(t\right)=-\frac{i}{\hbar}\left[\hat{H}\left(t\right),\hat{\sigma}\left(t\right)\right].\label{eq: LvN}
\end{equation}

In the basis of eigenstates of the dot and the lead sections of the
system the density matrix obtains the following block representation:

\begin{equation}
\hat{\sigma}\left(t\right)=\left(\begin{array}{cc}
\sigma_{d}\left(t\right) & \hat{\sigma}_{dL}\left(t\right)\\
\hat{\sigma}_{Ld}\left(t\right) & \hat{\sigma}_{L}\left(t\right)
\end{array}\right),\label{eq: density matrix general form-1}
\end{equation}

whose elements are given by $\sigma_{ij}(t)=\left\langle c_{i}(t)c_{j}^{\dagger}(t)\right\rangle $.
The corresponding block matrix representation of the Hamiltonian of
Eq. (\ref{eq: H of driven resonant level}) is given by: 

\begin{equation}
\hat{H}(t)=\left(\begin{array}{cc}
\varepsilon_{d}\left(t\right) & \hat{V}_{dL}\\
\hat{V}_{Ld} & \hat{H}_{L}
\end{array}\right).\label{eq: H}
\end{equation}

The initial diagonal density matrix represents a fully populated dot
($\sigma_{d}\left(t=0\right)=1$) and an empty lead ($\hat{\sigma}_{L}\left(t=0\right)=\hat{0}$;
In practice, we initially populate the lead levels according to a
Fermi-Dirac distribution, whose chemical potential and temperature
are set to $\mu=-50\hbar\gamma$ and $k_{B}\text{\ensuremath{T}}=0.25\gamma$,
respectively). By monitoring the diagonal element of the density matrix
that corresponds to the dot level we can follow its depopulation into
the lead levels. The resulting dynamics, which is represented by the
red curve in Fig. \ref{fig: pointcare}, captures well the short-time
($\gamma t<\gamma\left(h\Delta\varepsilon^{-1}\right)=20\pi$, reflecting
the highest frequency of the lead dynamics) exponential decay predicted
by the analytical treatment. However, at longer timescales, characteristic
Poincaré recurrences occur, reflecting the discrete nature of the
quasi-continuum representation of the lead or, equivalently, the reflection
of the scattered electron wavefunction from the far edge of the finite
lead model \cite{hod_driven_2016,ke_time-dependent_2010,koentopp_density_2008,kurth_time-dependent_2005,ochoa_energy_2016,zelovich_parameter-free_2017}.
Therefore, similar to previous multi-lead microcanonical transport
calculations \cite{di_ventra_transport_2004,ercan_tight-binding_2010,zelovich_state_2014},
it becomes evident that, while microcanonical simulations are not
limited to the WBA, the finite closed system model can mimic the behavior
of its open counterpart only for times shorter than the typical reflection
time-scale.

\subsection{Driven Liouville-von Neumann Simulations\label{subsec:The-driven-Liouville-von-Neuman}}

As mentioned above, the recently developed DLvN approach can eliminate
this limitation by expanding the capabilities of the microcanonical
approach to simulate truly open quantum systems. Similar to previous
multi-lead implementations of the DLvN approach\cite{zelovich_state_2014,chen_simple_2014,zelovich_molecule-lead_2015,zelovich_driven_2016,hod_driven_2016,zelovich_parameter-free_2017,morzan_electron_2017},
the LvN equation of motion for the single-lead setup considered herein
is augmented by sink and source terms that absorb outgoing electrons
(thus avoiding reflections) and inject thermalized electrons near
the system boundaries, respectively, as follows:

\begin{align}
\frac{\text{d}}{\text{d}t}\hat{\sigma}\left(t\right) & =-\frac{i}{\hbar}\left[\hat{H}\left(t\right),\hat{\sigma}\left(t\right)\right]-\Gamma\cdot\left(\begin{array}{cc}
0 & \frac{1}{2}\hat{\sigma}_{d,L}\left(t\right)\\
\frac{1}{2}\hat{\sigma}_{Ld}\left(t\right) & \hat{\sigma}_{L}\left(t\right)-\hat{\sigma}_{L}^{0}
\end{array}\right).\label{eq: DLvN}
\end{align}

The last term in Eq. \ref{eq: DLvN} serves to drive the lead section
towards a target equilibrium state of the form $\sigma_{ll'}^{0}=\delta_{ll'}f_{FD}\left(\varepsilon_{l},\mu,T\right)$,
where $f_{FD}\left(\varepsilon_{l};\mu,T\right)=\left[\exp\left(\left(\varepsilon_{l}-\mu\right)/\left(k_{B}T\right)\right)+1\right]^{-1}$
is a Fermi-Dirac equilibrium distribution with the chemical potential,
$\mu$, and electronic temperature, $T$, of the electronic reservoir,
to which the lead section is implicitly coupled, and $k_{B}$ is Boltzmann's
constant. The density matrix obtained from Eq. (\ref{eq: DLvN}) is
Hermitian, positive definite \cite{zelovich_driven_2016,elenewski_communication:_2017},
and normalized such that $\text{tr}\left[\hat{\sigma}\left(t\right)\right]=N_{tot}(t)$,
where $N_{tot}(t)$ is the instantaneous total number of electrons
in the system.

The driving rate, $\Gamma$, which can be extracted from the electronic
properties of the implicit reservoir \cite{zelovich_parameter-free_2017},
represents the timescale on which thermal relaxation takes place in
the lead, and is generally assumed to be fast relative to all other
processes of interest. If, however, the lead's driving rate is treated
as a phenomenological parameter, one should make sure that the relaxation
dynamics (i.e. the rate $\gamma$) of the dot itself is insensitive
to the choice of $\Gamma$. Since the latter broadens the lead levels,
the $\delta$-functions appearing in Eq. (\ref{eq: gm Fermi}) for
the dot's relaxation rate should be replaced by the corresponding
Lorentzian functions of the form $L_{l}\left(\varepsilon-\varepsilon_{l}\right)=\frac{1}{\pi}\frac{\hbar\Gamma/2}{\left(\varepsilon-\varepsilon_{l}\right)^{2}+\left(\hbar\Gamma/2\right)^{2}}$,
such that $\gamma(\varepsilon)=2\pi/\hbar\sum_{l}\left|V_{l}\right|^{2}L_{l}(\varepsilon-\varepsilon_{l})$.
If $\hbar\Gamma\gg\Delta\varepsilon$ the summand is a smooth function
of $\varepsilon_{l}$, which can be approximated by the following
integral:
\begin{equation}
\gamma(\varepsilon)\approx\frac{2\pi}{\hbar}\intop_{-W/2}^{W/2}\left|V\left(\varepsilon_{l}\right)\right|^{2}\rho\left(\varepsilon_{l}\right)L_{l}\left(\varepsilon-\varepsilon_{l}\right)d\varepsilon_{l}.
\end{equation}
Moreover, if $\left|V\left(\varepsilon_{l}\right)\right|^{2}$ and
$\rho\left(\varepsilon_{l}\right)$ do not (or only weakly) depend
on $\varepsilon_{l}$ (in practice, a softer requirement that $\left|V\left(\varepsilon_{l}\right)\right|^{2}\rho\left(\varepsilon_{l}\right)$
is independent of $\varepsilon_{l}$ is sufficient), they may be taken
out of the integral, and if furthermore $\hbar\Gamma\ll W$, the limits
of the remaining integral over $L_{l}\left(\varepsilon-\varepsilon_{l}\right)$
can be safely taken to infinity yielding a value of $1$ to a good
approximation. Hence, the wide band result of Eq. (\ref{eq: gm Fermi WBA})
stating that $\gamma\approx2\pi/\hbar\left|V\right|^{2}\rho$ is recovered
and the dynamics becomes independent of $\Gamma$. Thus, the lead
model should be chosen sufficiently large and its energy band should
be made sufficiently wide to allow for the value of $\Gamma$ to fulfill
the requirement $\Delta\varepsilon\ll\hbar\Gamma\ll W$ \cite{subotnik_nonequilibrium_2009,gruss_landauers_2016,elenewski_communication:_2017},
which assures that the finite lead model levels are sufficiently (but
not over-)broadened to mimic the continuous density-of-state within
the finite bandwidth of the corresponding bulk lead. We note in passing
that the above considerations are not just technical, and have been
repeatedly used to explain observations of molecular relaxation processes
involving isolated (on relevant timescales) large molecules \cite{bixon_intramolecular_1968},
where a discrete molecular spectrum appears (again on relevant timescales)
to act as a continuum \cite{bruch_quantum_2016,ludovico_dynamical_2014,ludovico_dynamics_2016,esposito_quantum_2015}\cite{carmeli_random_1982}.

To demonstrate the performance of this approach for the case of a
single-lead setup, we repeat the dot depopulation simulations of the
previous section with the same model Hamiltonian using the DLvN equation
of motion (\ref{eq: DLvN}) with a driving rate of $\hbar\Gamma=2\Delta\varepsilon=0.2\hbar\gamma$,
within the region spanned between $\Delta\varepsilon=0.1\hbar\gamma$
and $W=100\Delta\varepsilon=10\hbar\gamma$. The green curve in Fig.
\ref{fig: pointcare} presents the dot population as a function of
time obtained with the same initial conditions as those used in the
microcanonical simulation described above and a target equilibrium
lead density matrix of $\sigma_{ll'}^{0}=\delta_{ll'}f_{FD}\left(\varepsilon_{l},\mu=-50\hbar\gamma,T=0.25\hbar\gamma/k_{B}\right)$.
Clearly, the DLvN dynamics is able to reproduce both the short- and
the long-term analytical exponential decay of the dot population,
while eliminating the recurrences appearing in the microcanonical
simulations. Thus, it is shown that the DLvN effectively couples the
closed system to an external implicit bath resulting in a characteristic
open quantum system dynamics. Furthermore, when setting the implicit
bath's chemical potential at the center of the lead band and equal
to the dot's energy (both in the initial conditions and via the target
lead equilibrium density matrix, $\sigma_{ll'}^{0}=\delta_{ll'}f_{FD}\left(\varepsilon_{l},\mu=0,T=0.25\hbar\gamma/k_{B}\right)$),
the dot equilibrates to the expected half-filled state (see blue line
in Fig. \ref{fig: pointcare}). In order to verify that our results
are insensitive to the choice of driving rate we have repeated the
empty lead calculations for $\hbar\Gamma=0.2\Delta\varepsilon$ and
$\hbar\Gamma=20\Delta\varepsilon$. The comparison, presented in the
inset of Fig. \ref{fig: pointcare}, clearly demonstrate the weak
dependence of the simulated dynamics on the value of $\Gamma$. Notably,
this holds true also for the lower value chosen, which is outside
the validity range discussed above. This further demonstrates that,
with appropriate choice of model parameters, the DLvN approach can
effectively mimic different environmental conditions and may constitute
an effective numerical scheme to complement analytical treatments
in parameter regimes beyond their limiting assumptions. 

\section{Equilibrium Currents\label{sec: eq. currents}}

The simple single-lead model system discussed above demonstrated how
the DLvN approach can capture the \textit{total} current flowing between
the dot level and the lead manifold which, according to the analytical
treatment, is given by $J\left(t\right)=-dP_{d}\left(t\right)/dt=\gamma e^{-\gamma t}$
. However, in many applications, especially when evaluating thermodynamics
properties, it is useful to consider not only overall currents of
given observables but also their resolution with respect to other
observables. For example, the total current can be written in terms
of its energy resolved components as $J=\intop J\left(\varepsilon\right)d\varepsilon$,
where $J\left(\varepsilon\right)$ is the net particle flux leaving
the dot at a given energy $\varepsilon$. The latter can be then used
to evaluate thermodynamic quantities, such as the energy flux, $J_{E}=\intop\varepsilon J\left(\varepsilon\right)d\varepsilon$,
carried by the particles from the dot to the lead and the total heat
flux that they will produce in the environment when they eventually
get equilibrated in the lead, $J_{Q}=\intop d\varepsilon\left(\varepsilon-\mu\right)J\left(\varepsilon\right)$.

Such energy resolved currents can be evaluated via the numerical solution
of Eq. (\ref{eq: DLvN}), where the temporal variation of the occupation
of lead level $l$ is given by (see Appendix \ref{sec:Energy-Resolved-Currents}): 

\begin{equation}
\left(\frac{d\sigma_{ll}\left(t\right)}{dt}\right)=\frac{2}{\hbar}\Im\left(V_{ld}\sigma_{dl}\left(t\right)\right)-\Gamma\left(\sigma_{ll}\left(t\right)-\sigma_{ll}^{0}\right).\label{eq: flux into level l}
\end{equation}

We may now identify the first term on the right-hand-side of Eq. \ref{eq: flux into level l}
as the incoming particle flux from the dot into lead level $l$ and
the second term as the corresponding outgoing flux into the implicit
bath. Neglecting the broadening of the lead levels due to their coupling
to the bath and to the dot, the former can be used to evaluate the
energy resolved particle currents leaving the dot towards the lead
at energy $\varepsilon_{l}$:
\begin{equation}
J_{dL}\left(\varepsilon_{l},t\right)=\frac{2}{\hbar}\Im\left(V_{ld}\sigma_{dl}\left(t\right)\right),\label{eq: current at dot/lead interface}
\end{equation}

and the latter approximates the particle flux leaving the lead into
the implicit bath at the same energy:

\begin{equation}
J_{LB}\left(\varepsilon_{l},t\right)=\Gamma\left(\sigma_{ll}\left(t\right)-\sigma_{ll}^{0}\right).\label{eq: current at lead/bath}
\end{equation}

\begin{figure}
\begin{centering}
\includegraphics[scale=0.65]{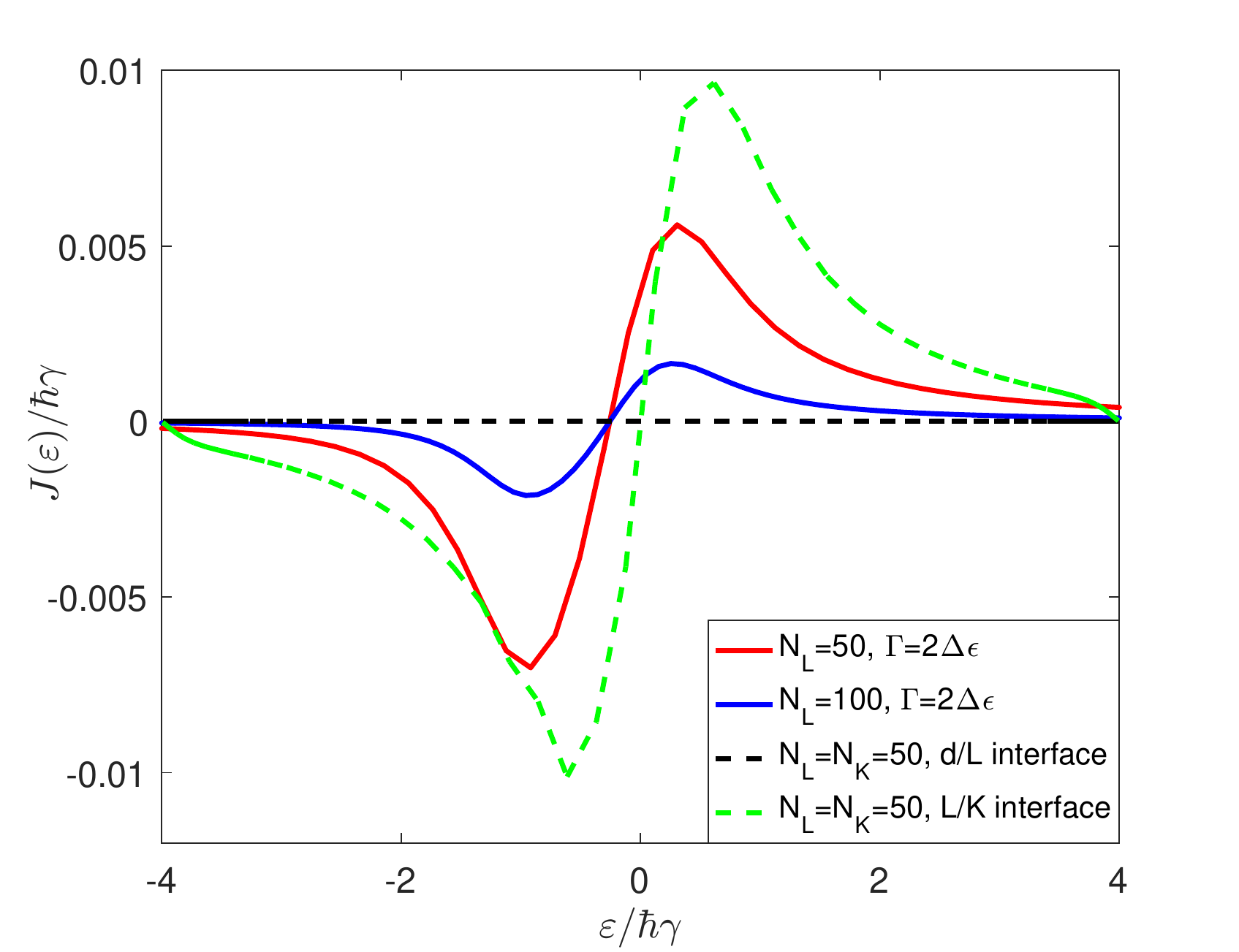}
\par\end{centering}
\caption{Spurious energy resolved equilibrium particle currents flowing between
the dot and the various levels of a driven lead of $N_{L}=50$ (full
red line) and $N_{L}=100$ (full blue line) levels, to which it is
directly coupled. For a system composed of a dot, a lead of $N_{L}=50$
levels, and a driven-lead of $N_{K}=50$ levels the equilibrium energy
resolved currents at the interface between the dot and the lead vanish
(dashed black line), while those at the interface between the lead
and the driven lead remain (dashed green line). \label{fig: eq_currents} }
\end{figure}

At equilibrium, we expect the total current and all of its energy
resolved components to vanish. Nevertheless, within the DLvN approach,
only the lead sections are directly equilibrated with their respective
implicit baths. This is essential for simulating non-equilibrium scenarios.
Therefore, since equilibration is not performed in the diagonal basis
of the entire finite model system and the dot section is not explicitly
equilibrated, for any finite lead model equilibrium can only be reached
approximately. As a result, when setting $d\hat{\sigma}/dt=0$ in
Eq. (\ref{eq: DLvN}) for the single-lead setup considered herein
(this can be readily done by solving a Sylvester equation \cite{subotnik_nonequilibrium_2009}
as detailed in Appendix \ref{sec:Sylvester-equation-for}), non-zero
dot-lead coherences $\sigma_{d,l}$ appear in the density matrix that
lead to spurious non-vanishing energy resolved equilibrium currents.
This, is clearly manifested by the red curve in Fig. \ref{fig: eq_currents}
showing the equilibrium energy resolved currents obtained for a lead
model consisting of $50$ equispaced levels that are spanning a bandwidth
of $W=10\hbar\gamma$. The lead levels, which are coupled to the dot
via $V=\sqrt{\frac{\hbar}{2\pi}\frac{\gamma}{\rho}}=\frac{\hbar\gamma}{\sqrt{10\pi}}$,
are driven at a rate of $\Gamma=0.4\gamma$ towards an equilibrium
Fermi Dirac population with chemical potential of $\mu=0$ and electronic
temperature of $k_{B}T=0.25\hbar\gamma$. The highest absolute current
value appears at the dot position of $\varepsilon_{d}=-\hbar\gamma$.
Notably, the total dot-lead current at equilibrium, obtained by summing
over all its energy resolved equilibrium components, $J_{dL}=\Sigma_{l}J^{eq}(\varepsilon_{l})$,
vanishes as expected. Nevertheless, the appearance of spurious non-vanishing
energy resolved equilibrium currents jeopardizes their validity for
calculating non-equilibrium thermodynamic properties, such as energy
and heat fluxes.

One remedy for this problem can be found by increasing the finite
lead model size. At the limit of an infinite lead, its spectrum mimics
well that of the entire dot+lead system and the effect of not directly
equilibrating the single dot level with the implicit bath becomes
negligible. This is demonstrated in Fig. \ref{fig: eq_currents},
where doubling the number of lead levels from $50$ (red) to $100$
(blue), while keeping the bandwidth at $W=10\hbar\gamma$ (yielding
$\Delta\varepsilon=\rho^{-1}=0.1\hbar\gamma$, $V=\frac{\hbar\gamma}{\sqrt{20\pi}}$,
and $\Gamma=2\Delta\varepsilon/\hbar=0.2\gamma$), reduces the magnitude
of the energy-resolved equilibrium currents. The effects of any residual
artificial currents on the calculation of non-equilibrium thermodynamic
properties can be eliminated by subtracting their equilibrium contribution
from the calculated dynamic properties.

\begin{figure}
\begin{centering}
\includegraphics[scale=0.4]{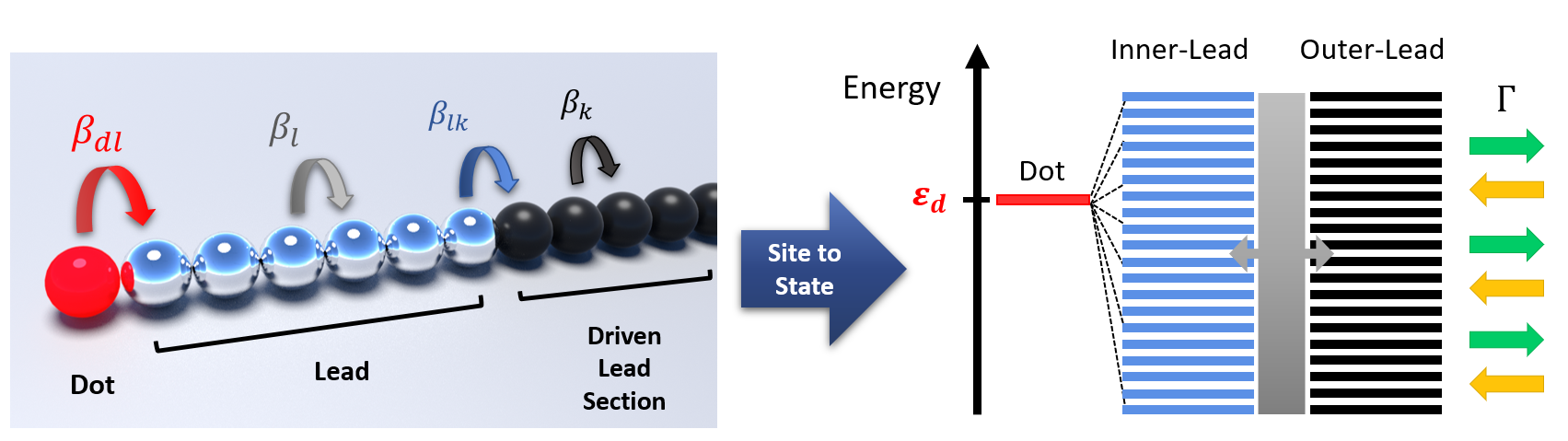}
\par\end{centering}
\caption{Site-to-state transformation. Left: Schematic site representation
of the tight-binding model for a one dimensional chain composed of
a dot (red), a lead (silver), and a driven-lead section (black). $\beta_{l}$
and $\beta_{k}$ denote the hopping integrals within the lead and
the driven-lead sections, respectively. $\beta_{dl}$ and $\beta_{lk}$
are the coupling matrix elements between the dot and the leftmost
site of the lead and between the rightmost site of the lead and the
leftmost site of the driven-lead section, respectively. Right: Scheme
of the single-particle state representation, where the dot level is
uniformly coupled to the eigenstates of the lead section that are
separately coupled to the manifold of eigenstates of the driven-lead
section that, in turn, are equilibrated at a rate $\Gamma$ with the
implicit external bath.\label{fig: site to state}}
\end{figure}

Alternatively, an ``extended-molecule'' strategy can be adopted.
Here, the system is divided into three (rather than two as before)
sections including (see left panel of Fig. \ref{fig: site to state}):
(i) the dot ($d$); (ii) the lead section adjacent to the dot that
is directly coupled to it ($L$); and (iii) a driven lead section
($K$). The first two constitute the extended-dot section that is
not directly coupled to the implicit bath. This buffers the dot from
the effects of the open boundary conditions that are imposed only
on the remote driven lead section. All physical quantities of interest
can now be evaluated from the dynamics of the dot section and the
dot/lead interface, where energy-resolved equilibrium currents vanish.
To demonstrate this, we consider a tight-binding chain consisting
of $N_{tot}=101$ sites, where the leftmost site serves as the dot,
the $N_{L}=50$ sites adjacent to the dot form the lead section, and
the remaining $N_{K}=50$ sites constitute the driven lead section.
The onsite energies of the dot, the lead, and the driven lead sites
are taken to be $\alpha_{d}=\alpha_{l}=\alpha_{k}=0$ eV, respectively.
The hopping integrals between the various lead sites $\left(\beta_{l}\right)$,
between the rightmost lead site and leftmost driven lead site $\left(\beta_{lk}\right)$,
and between the various driven lead sites $\left(\beta_{k}\right)$,
are set to $\beta_{l}=\beta_{lk}=\beta_{k}=0.2$ eV, respectively.
A weaker coupling of $\beta_{ld}=0.08$ eV is chosen between the dot
and the leftmost site of the lead section and all other hopping integrals
are nullified. This yields a (driven-)lead bandwidth of $W_{(driven-)lead}=4\beta_{(k)l}=0.8$
eV. The real-space tight-binding Hamiltonian matrix representation
of the system can be written in block matrix form as follows:

\begin{align}
\hat{H}\left(t\right)=\left(\begin{array}{ccc}
\varepsilon_{d}\left(t\right) & \hat{V}_{dL} & \hat{0}\\
\hat{V}_{Ld} & \hat{H}_{L} & \hat{V}_{LK}\\
\hat{0} & \hat{V}_{KL} & \hat{H}_{K}
\end{array}\right) & .\label{eq: H for dLK}
\end{align}

Here, the non-zero blocks are 

\begin{align}
 & \varepsilon_{d}\left(t\right)=\alpha_{d};\hat{V}_{dL}=\hat{V}_{Ld}^{\dagger}=\left(\begin{array}{ccc}
\beta_{ld} & 0 & \cdots\end{array}\right);\hat{H}_{L\left(K\right)}=\left(\begin{array}{ccc}
\alpha_{l\left(k\right)} & \beta_{l\left(k\right)} & 0\\
\beta_{l\left(k\right)} & \alpha_{l\left(k\right)} & \ddots\\
0 & \ddots & \ddots
\end{array}\right);\nonumber \\
 & \text{and }\hat{V}_{LK}=\hat{V}_{KL}^{\dagger}=\left(\begin{array}{ccc}
\vdots & \vdots & \iddots\\
0 & 0 & \cdots\\
\beta_{lk} & 0 & \cdots
\end{array}\right).\label{eq: H blokes of dLK}
\end{align}

Giving dimensions of $1\times N_{L}$ for $\hat{V}_{dL}=\hat{V}_{Ld}^{\dagger}$,
$N_{L(K)}\times N_{L(K)}$ for $\hat{H}_{L\left(K\right)}$, and $N_{L}\times N_{K}$
for $\hat{V}_{LK}=\hat{V}_{KL}^{\dagger}$. In order to impose the
DLvN boundary conditions on the eigenstates of the driven lead the
following unitary ``site-to-state'' transformation is performed
(see Fig. \ref{fig: site to state}) \cite{zelovich_state_2014}:

\begin{align}
\hat{U}=\left(\begin{array}{ccc}
1 & \hat{0} & \hat{0}\\
\hat{0} & \hat{U}_{L} & \hat{0}\\
\hat{0} & \hat{0} & \hat{U}_{K}
\end{array}\right) & ,\label{eq: U of dLK}
\end{align}

where $\hat{U}_{L}$ and $\hat{U}_{K}$ are the unitary matrices that
diagonalize $\hat{H}_{L}$ and $\hat{H}_{K}$, respectively, such
that $\tilde{\hat{H}}_{L/K}=\hat{U}_{L/K}^{\dagger}\hat{H}_{L/K}\hat{U}_{L/K}=\text{diag}\left\{ \varepsilon_{L/K}\right\} .$
The transformed Hamiltonian matrix has the same block structure as
its real-space counterpart:

\begin{align}
\hat{\tilde{H}}\left(t\right)=\hat{U}^{\dagger}\hat{H}\left(t\right)\hat{U}=\left(\begin{array}{ccc}
\varepsilon_{d}\left(t\right) & \tilde{\hat{V}}_{dL} & \hat{0}\\
\tilde{\hat{V}}_{Ld} & \tilde{\hat{H}}_{L} & \tilde{\hat{V}}_{LK}\\
\hat{0} & \tilde{\hat{V}}_{KL} & \tilde{\hat{H}}_{K}
\end{array}\right) & ,\label{eq: H diag of dLK}
\end{align}

where $\hat{\tilde{V}}_{dL}=\hat{\tilde{V}}_{Ld}^{\dagger}$ hold
the couplings between the dot and the various lead levels and $\hat{\tilde{V}}_{LK}=\hat{\tilde{V}}_{KL}^{\dagger}$
store the couplings between the latter and the different driven lead
levels (see right panel of Fig. \ref{fig: site to state}). In order
to mimic the simulation conditions used above, where the dot is uniformly
coupled to all lead levels, we replace all elements in $\hat{\tilde{V}}_{dL}$
(and $\hat{\tilde{V}}_{Ld}^{\dagger}$) by their highest value of
$V\simeq0.0158$ eV constituting the maximum of the corresponding
Newns Anderson coupling band \cite{zelovich_molecule-lead_2015}.
Given the density of lead states, $\rho=50/\left(4\beta_{l}\right)=62.5$
eV$^{-1}$, this yields $\hbar\gamma=0.0985$ eV (see Eq. \ref{eq: gm Fermi WBA}
above), which is comparable to the value of $0.1$ eV used above.

The resulting DLvN equation of motion, written in the basis of eigenstates
of the dot, lead, and driven lead sections, has the form:

\begin{align}
\frac{\text{d}}{\text{d}t}\hat{\sigma}\left(t\right) & =-\frac{i}{\hbar}\left[\hat{\tilde{H}}\left(t\right),\hat{\sigma}\left(t\right)\right]-\Gamma\cdot\left(\begin{array}{ccc}
0 & \hat{0} & \frac{1}{2}\hat{\sigma}_{dK}\left(t\right)\\
\hat{0} & \hat{0} & \frac{1}{2}\hat{\sigma}_{LK}\left(t\right)\\
\frac{1}{2}\hat{\sigma}_{Kd}\left(t\right) & \frac{1}{2}\hat{\sigma}_{KL}\left(t\right) & \hat{\sigma}_{K}\left(t\right)-\hat{\sigma}_{K}^{0}
\end{array}\right),\label{eq: DLvN for dLK}
\end{align}

where $\hat{\sigma}_{K}^{0}$ is the target equilibrium density matrix
imposed by the implicit bath on the driven-lead section with $\mu=0$
and $k_{B}T=0.25\hbar\gamma$ and the driving rate is chosen as $\Gamma=2/\left(\hbar\rho\right)=0.0486$
fs$^{\ensuremath{-1}}$. When setting $d\hat{\sigma}/dt=0$ (see Appendix
\ref{sec:Sylvester-equation-for}), the energy resolved currents between
the dot and the lead section now vanish (see dashed black curve in
Fig. \ref{fig: eq_currents}) as required. Nevertheless, the equilibrium
state of the entire finite system remains approximate and the spurious
currents have been just driven away toward the (less physically relevant)
interface between the lead and the driven lead sections. This is demonstrated
by the dashed green curve in Fig. \ref{fig: eq_currents}, where we
plot the total current flowing from the lead section to the various
driven lead levels, k, calculated via 

\begin{equation}
J_{LK}\left(\varepsilon_{k}\right)=\frac{2}{\hbar}\sum_{l}^{N_{L}}\left(\Im\left(H_{kl}\sigma_{lk}\left(t\right)\right)\right).
\end{equation}

This, therefore, clearly demonstrates that care should be exercised
when utilizing numerical schemes using finite models to simulate (thermo)dynamic
properties of open quantum systems. Brute force application of such
schemes may lead to unphysical results that are strongly influenced
by the applied boundary conditions.

\section{Finite Bandwidth Effects\label{subsec:  Finite Bandwidth Effects}}

The numerical examples provided above considered a static dot level
placed sufficiently far from the lead's band-edges and situated symmetrically
between them. In non-equilibrium thermodynamic calculations, however,
we will often be interested in simulating time-dependent perturbations
applied to the system. These may include time-dependent external fields
or varying gate potentials that dynamically shift the dot's level
energy. In such cases, it may become inevitable to position the dot
level in the vicinity of the lead's band edges. Hence, it is important
to understand both the physical and the numerical implications of
approaching the band-edges of the modeled environment. This is especially
true in the context of comparisons with, and extensions of, analytical
treatments that, as stated above, often make simplifying assumptions,
such as the WBA (see Sec. \ref{sec: eq. currents} above) that treats
the environment as an infinite energy band of uniform and continuous
density-of-states.

To demonstrate this, we study the changes in equilibrium total number
of particles and electronic energy of the finite system upon shifting
the dot away from the band center towards the upper band edge. Comparing
the results for increasing band-widths to the predictions of the analytical
WBA treatment allows us to assess the importance of band-edge effects
and the convergence of the numerical model to the WBL. To this end,
we consider the isolated system consisting of the dot level and a
finite lead manifold of $N_{l}$ states. We choose a uniform density
of lead levels of $\rho=10(\hbar\gamma)^{-1}$ that are uniformly
coupled to the dot level via $V=\sqrt{\frac{\hbar\gamma}{2\pi\rho}}=\frac{\hbar\gamma}{\sqrt{20\pi}}$
(see Eq. \ref{eq: gm Fermi}). The dot level is first positioned at
the center of the lead levels band, $\varepsilon_{d}=0$, and the
Hamiltonian matrix of the entire closed system is diagonalized. The
eigenstates are then occupied according to the Fermi-Dirac distribution
and the equilibrium number of particles and total electronic energy
are calculated as: 
\begin{equation}
N(\varepsilon_{d}=0,W)=\sum_{j=0}^{N_{l}+1}f_{FD}\left(\varepsilon_{j};\mu=0,T=0.25\hbar\gamma/k_{B}\right)\label{eq:Equilibrium Total Occupation}
\end{equation}
 and 
\begin{equation}
E(\varepsilon_{d}=0,W)=\sum_{j=0}^{N_{l}+1}f_{FD}\left(\varepsilon_{j};\mu=0,T=0.25\hbar\gamma/k_{B}\right)\varepsilon_{j},\label{eq:Equilibrium Total Energy}
\end{equation}
respectively, where $W=N_{l}/\rho$. Similarly, we obtain $N(\varepsilon_{d}=2\hbar\gamma,W)$
and $E(\varepsilon_{d}=2\hbar\gamma,W)$ by positioning the dot level
$2\hbar\gamma$ above the lead's band center, and calculate the variations
$\Delta N_{num}(W)=N\left(\varepsilon_{d}=2\hbar\gamma,W\right)-N\left(\varepsilon_{d}=0,W\right)$
and $\Delta E_{num}(W)=E\left(\varepsilon_{d}=2\hbar\gamma,W\right)-E\left(\varepsilon_{d}=0,W\right)$.
To assess the correspondence between the numerical calculation and
the analytical WBA results we repeat this procedure for increasing
lead's bandwidth by increasing the number of lead states while keeping
their density fixed. At the limit of infinite bandwidth we expect
the numerical results to converge to the analytical WBA values of:
\begin{equation}
\Delta N_{analytic}=\intop_{-\infty}^{\infty}\left[A\left(\varepsilon-2\hbar\gamma;\Delta=0;\gamma\right)-A\left(\varepsilon-0;\Delta=0;\gamma\right)\right]f_{FD}\left(\varepsilon;\mu=0,T=0.25\hbar\gamma/k_{B}\right)d\varepsilon
\end{equation}
 and 
\begin{equation}
\Delta E_{analytic}=\intop_{-\infty}^{\infty}\left[A\left(\varepsilon-2\hbar\gamma;\Delta=0;\gamma\right)-A\left(\varepsilon-0;\Delta=0;\gamma\right)\right]f_{FD}\left(\varepsilon;\mu=0,T=0.25\hbar\gamma/k_{B}\right)\varepsilon d\varepsilon.
\end{equation}

\begin{figure}
\begin{centering}
\includegraphics[scale=0.75]{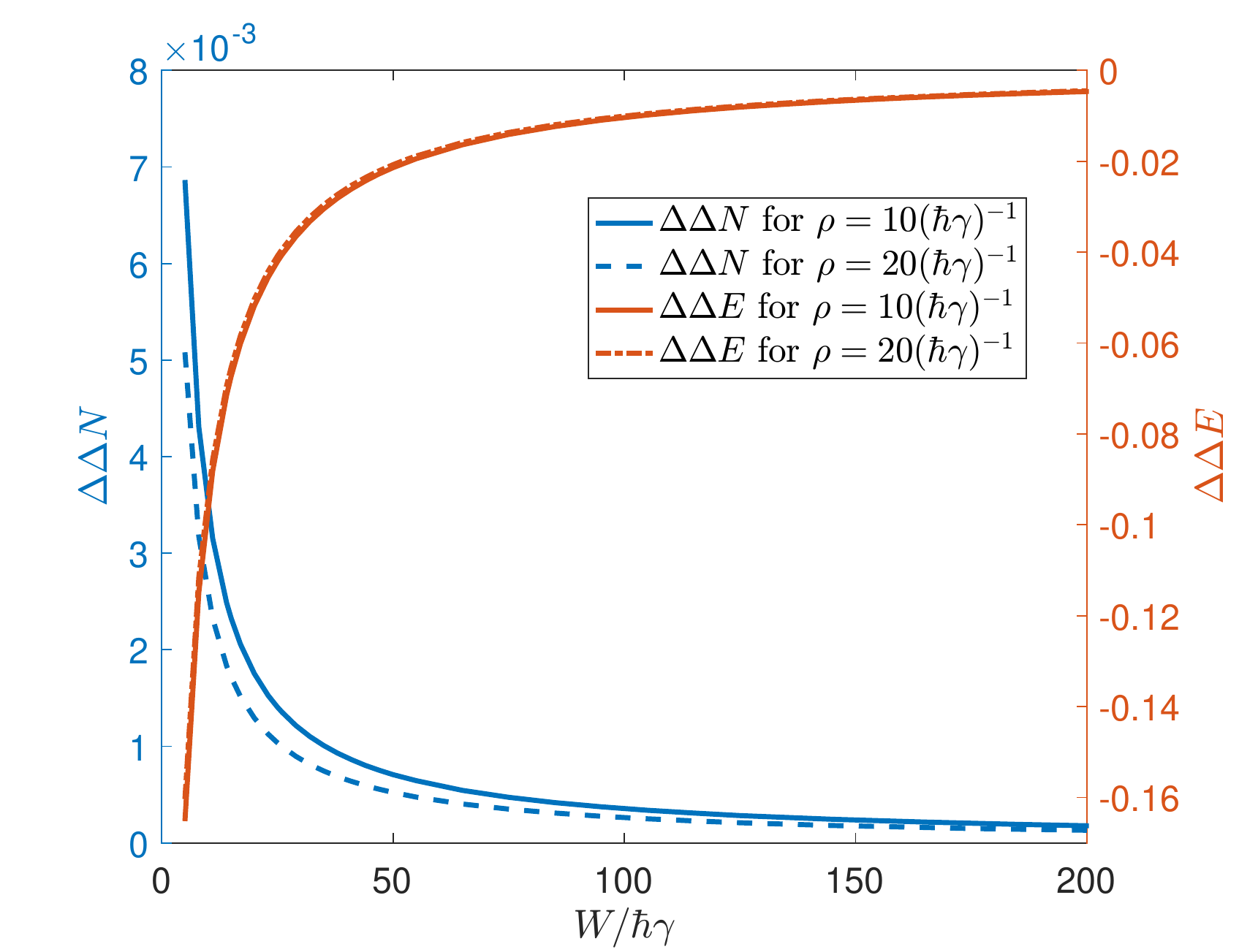}
\par\end{centering}
\caption{Convergence of the calculated equilibrium occupation, $\Delta\Delta N$
(blue), and electronic energy, $\Delta\Delta E$ (red), variations
of the finite lead model system towards the wide band limit. The results
are obtained for lead density of lead states of $\rho=10(\hbar\gamma)^{-1}$
(full lines) and $\rho=20(\hbar\gamma)^{-1}$ (dashed lines). \label{fig: integration limits} }
\end{figure}

In practice, we calculate these integrals numerically with integration
bounds of $W=3000\hbar\gamma$, such that increasing the bounds to
$W=3500\hbar\gamma$ gives a difference of $0.02\%$ for the energy
and $9\times10^{-6}\%$ for the particle number. Note that when comparing
the numerical results to the analytical values, the lead levels occupations
are assumed to be insensitive to the dot level position such that
$\Delta N_{num}$ and $\Delta E_{num}$ reflect only the change in
dot occupation and energy contribution, like their analytical counterparts.
Fig. \ref{fig: integration limits} shows the relative deviation of
the change of number of particles (full blue line) with respect to
the analytical WBA result $\Delta\Delta N(W)=[\Delta N_{num}(W)-\Delta N_{analytic}]/\Delta N_{analytic}$
and the corresponding relative energy deviation (full red line) $\Delta\Delta E(W)=[\Delta E_{num}(W)-\Delta E_{analytic}]/\Delta E_{analytic}$
as a function of bandwidth, $W$. We find that, for a finite band
model, the change in number of particles upon the upshift of the dot
level from the band center is larger than the analytical WBA result,
whereas the corresponding change in electronic energy is smaller than
its WBA counterpart. As expected, both $\Delta N_{num}$ and $\Delta E_{num}$
converge to the corresponding analytical WBL values with increasing
finite lead model bandwidth. Notably, the particle number change converges
faster than the electronic energy change, such that at a bandwidth
of $200\hbar\gamma$ the deviation of $\Delta N_{num}$ from $\Delta N_{analytic}$
reduces to $0.02\%$ while the corresponding deviation in the electronic
energy is still larger than $0.5\%$. To rationalize this observation
we note that the integrand of $\Delta E_{analytic}$ includes $\varepsilon$
itself, which diverges at the integration limits, and hence slows
the convergence of the integrand at any finite integration range.
This exemplifies a general behavior that different observables converge
at a different rate with system parameters, thus care should be taken
to separately converge them. To further verify that these result are
converged with respect to the choice of density of lead levels we
have repeated the calculations for a density of $\rho=20(\hbar\gamma)^{-1}$
obtaining only minor deviations for the particle number and energy
variations (see dashed blue and red lines in Fig. \ref{fig: integration limits}).
The analysis presented above thus demonstrates that numerical treatments
can simulate various environment models ranging from simplistic wide-band
baths to more complex finite-band baths that are not restricted to
uniform density-of-states and/or system bath couplings.

\begin{figure}[H]
\begin{centering}
\includegraphics[scale=0.75]{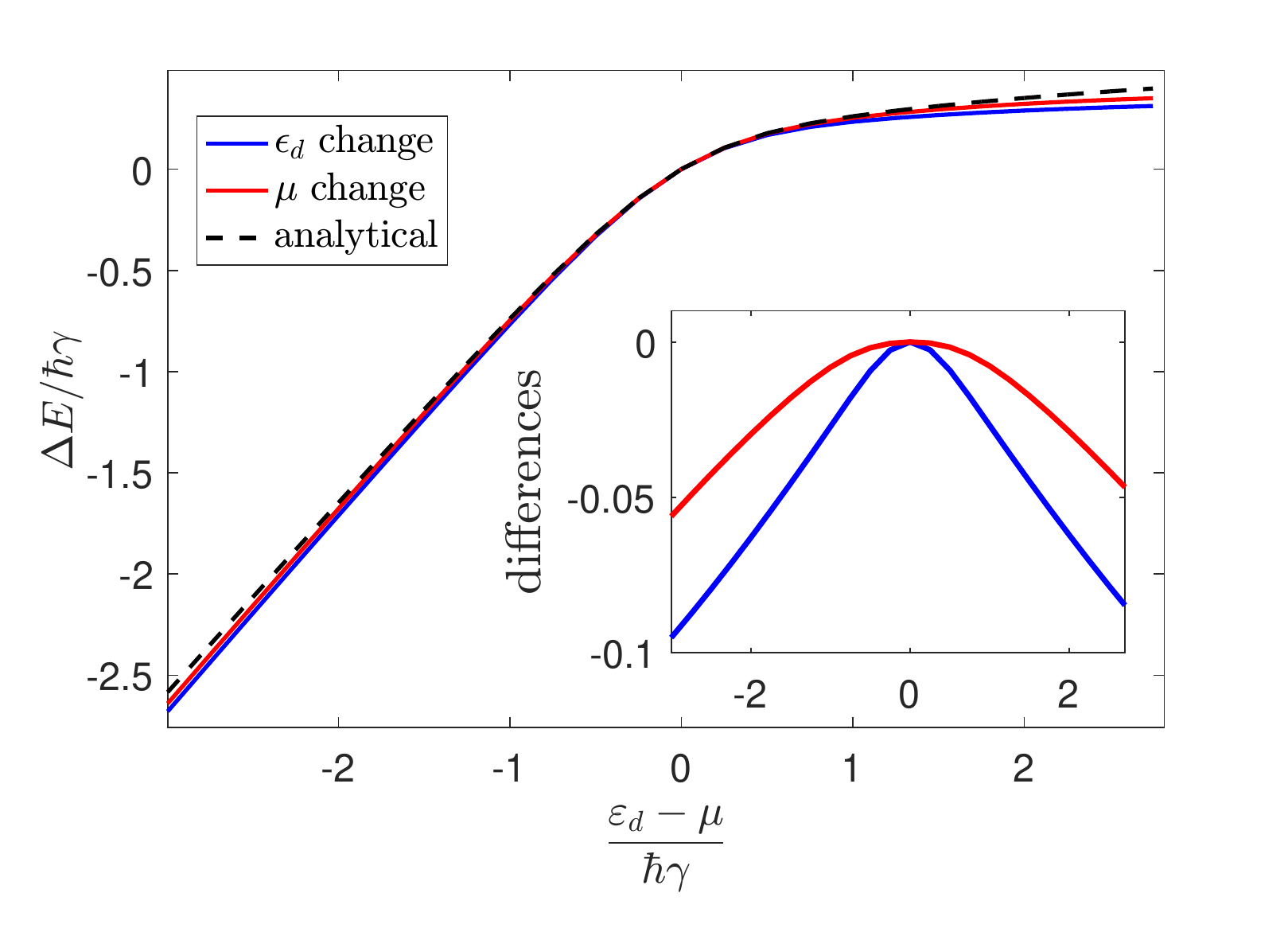}
\par\end{centering}
\caption{Comparison between the numerical evaluation (full blue and red lines)
of the contribution to the total equilibrium electronic energy of
a dot that is uniformly coupled to a discrete set of lead levels of
finite bandwidth and the corresponding analytical WBA result (dashed
black line). The numerical evaluation is performed either by shifting
the dot's position while keeping the chemical potential fixed at the
lead band center (blue) or vice versa (red). In all graphs, the Y-axis
origin is set to the dot's contribution to the total energy when placed,
along with the chemical potential, at the center of the lead's band.
\textbf{Inset: }The differences between the numerical and the analytical
evaluations of the dot's energy contribution to the total equilibrium
electronic energy as a function of its position along the leads band.
The line colors correspond to those in the main panel.\label{fig: mu vs. eps_d change} }
\end{figure}

The same holds true not only for simulating complex bath models but
also for studying dynamical processes of the system itself. In the
resonant level model discussed herein the latter may translate to
dynamical shifts of the dot's level energy across the lead's band.
Nevertheless, prior to performing dynamical simulations one should
first verify that the numerical approach can reproduce quasi-static
results. To this end, we repeated the procedure detailed above using
a finite lead model consisting of $500$ levels spanning a bandwidth
of $W=10\hbar\gamma$, which, according to Fig. \ref{fig: integration limits},
reproduces WBA occupations and energetics down to $\sim0.4\%$ and
$\sim9.5\%$, respectively. The dot level is then uniformly coupled
to all lead states with $V=\frac{\hbar\gamma}{\sqrt{100\pi}}$ and
its energy is varied around the chemical potential of the lead. For
each dot level position, $\varepsilon_{d}$, the Hamiltonian of the
entire finite model system is diagonalized and its eigenstates, $\{\left|j\right\rangle \}$,
are occupied according to the equilibrium Fermi-Dirac distribution.
As our observable we choose the dot section contribution to the total
electronic energy of the system. In the above treatment we have assumed
that the lead section populations are insensitive to the dot level
position, such that any change in total energy of the system reflects
only the dot's contribution. Alternatively, we can evaluate it in
the eigenbasis of the entire system via $E_{d}^{num}\left(\varepsilon_{d}\right)=\sum_{j}\varepsilon_{j}f(\varepsilon_{j})\left|\left\langle d|j\right\rangle \right|^{2}$,
where the sum runs over all eigenstates, and their individual contributions
to the total electronic energy $\varepsilon_{j}f(\varepsilon_{j})$
($\varepsilon_{j}$ and $f(\varepsilon_{j})$ being the orbital energy
and equilibrium occupation, respectively) are scaled by their projection
on the dot section, $\left|\left\langle d|j\right\rangle \right|^{2}$.
In Fig. \ref{fig: mu vs. eps_d change} we compare the numerical value
(full blue line) obtained for $\Delta E_{d}^{num}\left(\varepsilon_{d}\right)=E_{d}^{num}\left(\varepsilon_{d}\right)-E_{d}^{num}\left(\mu=0\right)$
at various dot level positions in a range of $\pm2\hbar\gamma$ around
the chemical potential (which is kept fix at $\mu=0$) and an electronic
temperature of $T=0.25\hbar\gamma/k_{B}$, to the analytical WBA results
(dashed black line) obtained, as above, from:
\begin{equation}
\Delta E_{d}^{analytic}\left(\varepsilon_{d}\right)=\intop_{-\infty}^{\infty}\left[A\left(\varepsilon-\varepsilon_{d};\Delta=0;\gamma\right)-A\left(\varepsilon-\mu;\Delta=0;\gamma\right)\right]f_{FD}\left(\varepsilon;\mu=0,T=0.25\hbar\gamma/k_{B}\right)\varepsilon d\varepsilon.\label{eq: E_analytic}
\end{equation}
As noted above, in practice we calculate these integrals numerically
with integration bounds of $W=3000\hbar\gamma$. In the vicinity of
the lead's Fermi energy the agreement between the two calculations
is seen to be excellent. Minor deviations between the two develop
as the dot position approaches the band edges of the finite lead model
(see inset of Fig. \ref{fig: mu vs. eps_d change}). To avoid such
finite-bandwidth effects and achieve better agreement between the
numerical results and the analytical wide band approximation we suggest
an alternative approach, where the dot level is kept fixed at the
lead's band center (symmetrically between the two band edges) and
the chemical potential of the lead is varied around it. The results
of this practice are presented by the full red line in Fig. \ref{fig: mu vs. eps_d change}
showing better agreement with the analytical WBA results as is clearly
demonstrated in the inset.

Having established that our method can reproduce quasi-static results
we can now turn to discuss dynamic variations of the dot's level position.
Simulating such processes with the closed system treatment presented
above will require extremely large lead models to prevent backscattering
from the finite-model boundaries during the relevant simulation time-scales.
As demonstrated above, this can be readily avoided by using the DLvN
approach, where broadening of the discrete manifold of levels of a
relatively small finite lead model allows to mimic the continuous
density-of-state of a (semi-)infinite bath. Result of such simulations
can provide valuable dynamical information for finite band bath models.
However, one might also wish to use the numerical treatment to study
dynamical processes in the wide bath band limit, which cannot be accessed
using current analytical treatments. Here, as well, we could increase
the finite lead model until convergence with respect to its bandwidth
is obtained. Within the context of dynamical simulations, however,
this would considerably increase the computational burden and defeat
the main purpose of the DLvN approach. Hence, again, we offer an alternative
by assuming that the difference between the numerical finite-lead-band
result and the analytical WBA results depend weakly on the rate of
dot level shift. If this assumption is valid, we can extract this
difference from a quasi-static calculation, where both numerical and
analytical results are available: 
\begin{equation}
\delta E_{d}\left(\varepsilon_{d}\right)=E_{d}^{num,QS}(\varepsilon_{d})-E_{d}^{analytic,QS}\left(\varepsilon_{d}\right).\label{eq: delta =00003D analytic-num}
\end{equation}
By subtracting this difference from the dynamic finite-bandwidth model
numerical results we obtain an estimate of the corresponding WBL results.
To demonstrate this, we use a relatively small lead model consisting
of $N_{L}=100$ lead states spanning a bandwidth of $W=10\hbar\gamma$.
Notably, the latter is taken deliberately insufficient to achieve
convergence to the WBL (see Fig. \ref{fig: integration limits}).
We solve the DLvN equation of motion using a driving rate of $\Gamma=0.2\hbar\gamma$,
and a target density that provides a chemical potential of $\mu=0$
and an electronic temperature of $k_{B}T=0.25\gamma$, starting at
equilibrium with the dot level positioned at $\varepsilon_{d}=\mu-3\hbar\gamma$
and shifting it at a constant rate of $\dot{\varepsilon}_{d}/\left(\hbar\gamma^{2}\right)=0.6582$
up to $\varepsilon_{d}=\mu+3\hbar\gamma$. Using the density matrix
of the entire system we can evaluate the temporal evolution of the
dot's contribution to the total electronic energy using the following
projection: 
\begin{equation}
E_{d}^{num}\left(t\right)=\frac{1}{2}\left\langle d\right|\hat{H}\left(t\right)\hat{\sigma}\left(t\right)+\hat{\sigma}\left(t\right)\hat{H}\left(t\right)\left|d\right\rangle ,\label{eq: E_num}
\end{equation}
which is symmetrized to be real valued.

\begin{figure}
\begin{centering}
\includegraphics[scale=0.75]{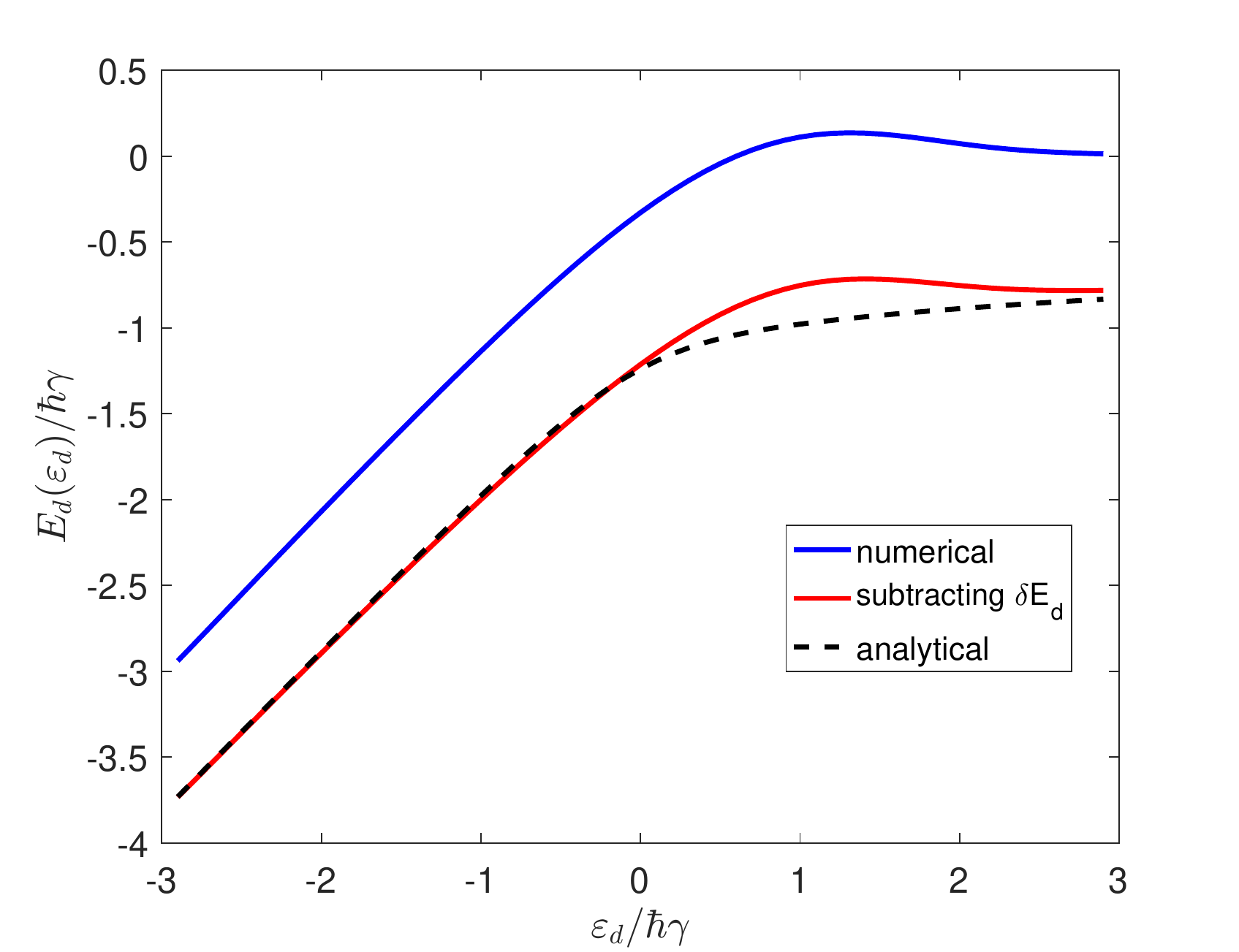}
\par\end{centering}
\caption{Dynamic contribution (full blue line) to the total electronic energy
due to a dot that is up-shifted at a finite rate across a discrete
set of lead levels of finite bandwidth, to which it is uniformly coupled.\textbf{
}Estimation of the corresponding WBL behavior, obtained by subtracting
$\delta E_{d}\left(\varepsilon_{d}\left(t\right)\right)$ from the
simulation results, is presented in red. The quasi-static analytical
WBA results are given as reference by the dashed black line. \label{fig: numerical corrected for finite BW}}
\end{figure}

The results of this calculation (full blue line in Fig. \ref{fig: numerical corrected for finite BW})
differ from the quasi-static analytical WBA results (full red line
in Fig. \ref{fig: numerical corrected for finite BW}) over the entire
range of dot positions studied. This can be attributed to two main
factors: (i) the finite lead bandwidth of the numerical model compared
to the infinite bath band assumed in the analytical case; and (ii)
the finite dot level shift rate used in the numerical simulation,
which pushes the system out of its equilibrium state that is assumed
by the quasi-static analytical treatment. Under the assumption mentioned
above, we can eliminate the effect of the former by adding $\delta E_{d}\left(\varepsilon_{d}\left(t\right)\right)=\delta E_{d}\left(\mu-3\hbar\gamma+\dot{\varepsilon}_{d}t\right)$
to the calculated $E_{d}^{num}\left(t\right)$. This allows us to
estimate effects of dynamical dot level shifts at the wide-band bath
limit. Comparing the red line in Fig. \ref{fig: numerical corrected for finite BW}
for $E_{d}^{num,WBA}\left(t\right)=E_{d}^{num}\left(t\right)-\delta E_{d}\left(\varepsilon_{d}\left(t\right)\right)$
to the dashed black line for $E_{d}^{analytic,QS}\left(\varepsilon_{d}\right)$
we see that, up to the lead's Fermi energy the dynamical result resemble
the quasi-static behavior, exhibiting a linear increase. This reflects
the fact that in this region the dot remains fully occupied and the
variation of the dot's contribution to the total electronic energy
stems only from changing its position. When approaching the Fermi
level, the dot gradually empties into the lead. Hence the energy rise
of $E_{d}^{num,WBA}\left(t\right)$ due to the upshift of $\varepsilon_{d}$
is countered by the dot's depopulation and its slope reduces. Noticeably,
when the rate of the dot level up-shift becomes comparable or larger
than $\gamma$, its emptying into the lead lags behind that of the
quasi-static case. This results in the rate-dependent hysteresis evident
in Fig. \ref{fig: numerical corrected for finite BW}, where $E_{d}^{num,WBA}\left(t\right)$
overshoots $E_{d}^{analytic,QS}\left(\varepsilon_{d}\right)$ in the
vicinity of the lead Fermi energy. The analysis presented above thus
demonstrates how DLvN based simulations can be used to study dynamical
effects in open quantum systems in a wide range of system and bath
parameters and extract important information relevant for evaluating
their non-equilibrium thermodynamic properties.

\section{Summary and Conclusions\label{sec:Summary and Conclusions}}

The study of non-equilibrium dynamics and thermodynamics of open quantum
systems is currently gaining increasing theoretical and experimental
interest. Simple analytical treatments provide valuable insights regarding
the extension of thermodynamic quantities towards non-equilibrium
conditions. These, however, are often based on simplifying assumptions
regarding the structure of the system, the environment, and their
inter-coupling, thus limiting their validity to specific parameter
ranges. Numerical approaches, such as the Driven Liouville-von Neumann
methodology, can help bridge the gap between phenomenological analytical
treatments and realistic experimental scenarios. In this paper, we
presented a brief outline of the DLvN approach and discussed some
important methodological aspects of its utilization for studying non-equilibrium
(thermo)dynamic properties. Specifically, we have demonstrated that
DLvN simulations using finite model systems can capture the depopulation
dynamics of an impurity electronic state uniformly coupled to an infinite
bath of continuous and constant density of states. We have shown that
when evaluating energy resolved quantities based on DLvN simulations,
care should be taken to avoid the effects of spurious equilibrium
currents resulting from the inherently approximate equilibrium state
imposed on the system. We have further studied the convergence of
DLvN numerical simulations towards the wide band bath limit upon increase
in the bandwidth of the finite lead model. Finally, we have shown
how one can obtain reliable static and dynamic wide band results using
relatively small model systems either via efficient cancellation of
finite-bandwidth effect or by their direct subtraction from the simulated
properties. Importantly, these methodological considerations may be
relevant to other numerical techniques for simulating electron dynamics
in open quantum systems. Hence, with the understanding gained herein,
numerical approaches, such as the DLvN methodology, may become efficient
tools for simulating non-equilibrium quantum thermodynamics in experimentally
relevant regimes that are out of the reach of current analytical treatments.

\section{Acknowledgments}

IO gratefully acknowledges the support of the Adams Fellowship Program
of the Israel Academy of Sciences and Humanities, and the Naomi Foundation
through the Tel-Aviv University GRTF Program. The research of AN is
supported by the U.S. National Science Foundation (Grant No. CHE1665291,the
Israel-U.S. Binational Science Foundation, the German Research Foundation
(DFG \emph{TH} 820/11-1), and the University of Pennsylvania. OH is
grateful for the generous financial support of the Israel Science
Foundation under Grant No. 1740/13 and the Center for Nanoscience
and Nanotechnology of Tel-Aviv University. 

\bibliographystyle{unsrt}
\bibliography{10C__Users_USER_Google_Drive_draft_of_Numerical___nd_relaxation_inbaloz_paper_folder_endnotes}

\begin{thebibliography}{10}

\bibitem{nitzan_electron_2003}
Abraham Nitzan and Mark~A. Ratner.
\newblock Electron {Transport} in {Molecular} {Wire} {Junctions}.
\newblock {\em Science}, 300(5624):1384--1389, 2003.

\bibitem{cuevas_molecular_2010}
Juan~Carlos Cuevas and Elke Scheer.
\newblock {\em Molecular {Electronics}: {An} {Introduction} to {Theory} and
  {Experiment}}.
\newblock World {Scientific} series in nanoscience and nanotechnology ; v. 1.
  World Scientific, 2010.

\bibitem{ghosh_nanoelectronics_2016}
Avik Ghosh.
\newblock {\em Nanoelectronics {A} {Molecular} {View}}.
\newblock World Scientific, 2016.

\bibitem{galperin_molecular_2012}
Michael Galperin and Abraham Nitzan.
\newblock Molecular optoelectronics: the interaction of molecular conduction
  junctions with light.
\newblock {\em Physical Chemistry Chemical Physics}, 14(26):9421, 2012.

\bibitem{bruch_quantum_2016}
Anton Bruch, Mark Thomas, Silvia Viola~Kusminskiy, Felix von Oppen, and Abraham
  Nitzan.
\newblock Quantum thermodynamics of the driven resonant level model.
\newblock {\em Physical Review B}, 93(11):115318, March 2016.

\bibitem{ludovico_dynamical_2014}
Mar{\'i}a~Florencia Ludovico, Jong~Soo Lim, Michael Moskalets, Liliana
  Arrachea, and David S{\'a}nchez.
\newblock Dynamical energy transfer in ac-driven quantum systems.
\newblock {\em Physical Review B}, 89(16):161306(R), April 2014.

\bibitem{ludovico_dynamics_2016}
Mar{\'i}a~Florencia Ludovico, Michael Moskalets, David S{\'a}nchez, and Liliana
  Arrachea.
\newblock Dynamics of energy transport and entropy production in ac-driven
  quantum electron systems.
\newblock {\em Physical Review B}, 94(3):035436, July 2016.

\bibitem{esposito_quantum_2015}
Massimiliano Esposito, Maicol~A. Ochoa, and Michael Galperin.
\newblock Quantum {Thermodynamics}: {A} {Nonequilibrium}
  {Green}{\textquoteright}s {Function} {Approach}.
\newblock {\em Physical Review Letters}, 114(8):080602, February 2015.

\bibitem{tuovinen_phononic_2016}
Riku Tuovinen, Niko S{\"a}kkinen, Daniel Karlsson, Gianluca Stefanucci, and
  Robert van Leeuwen.
\newblock Phononic heat transport in the transient regime: {An} analytic
  solution.
\newblock {\em Physical Review B}, 93(21):214301, June 2016.

\bibitem{gelbwaser-klimovsky_thermodynamics_2015}
David Gelbwaser-Klimovsky, Wolfgang Niedenzu, and Gershon Kurizki.
\newblock Thermodynamics of {Quantum} {Systems} {Under} {Dynamical} {Control}.
\newblock In {\em Advances {In} {Atomic}, {Molecular}, and {Optical}
  {Physics}}, volume~64, pages 329--407. Elsevier, 2015.

\bibitem{arrachea_microscopic_2012}
Liliana Arrachea, Eduardo~R. Mucciolo, Claudio Chamon, and Rodrigo~B. Capaz.
\newblock Microscopic model of a phononic refrigerator.
\newblock {\em Physical Review B}, 86(12):125424, September 2012.

\bibitem{kosloff_quantum_2014}
Ronnie Kosloff and Amikam Levy.
\newblock Quantum {Heat} {Engines} and {Refrigerators}: {Continuous} {Devices}.
\newblock {\em Annual Review of Physical Chemistry}, 65(1):365--393, April
  2014.

\bibitem{haug_quantum_2008}
Hartmut Haug and Antti-Pekka Jauho.
\newblock {\em Quantum {Kinetics} in {Transport} and {Optics} of
  {Semiconductors}}, volume 123.
\newblock Springer-Verlag Berlin Heidelberg, 2 edition, 2008.

\bibitem{chen_simple_2014}
Liping Chen, Thorsten Hansen, and Ignacio Franco.
\newblock Simple and {Accurate} {Method} for {Time}-{Dependent} {Transport}
  along {Nanoscale} {Junctions}.
\newblock {\em The Journal of Physical Chemistry C}, 118(34):20009--20017,
  August 2014.

\bibitem{cheng_simulating_2006}
Chiao-Lun Cheng, Jeremy~S. Evans, and Troy Van~Voorhis.
\newblock Simulating molecular conductance using real-time density functional
  theory.
\newblock {\em Physical Review B}, 74(15):155112, October 2006.

\bibitem{di_ventra_transport_2004}
M~Di~Ventra and T~N Todorov.
\newblock Transport in nanoscale systems: the microcanonical versus
  grand-canonical picture.
\newblock {\em Journal of Physics: Condensed Matter}, 16(45):8025--8034,
  November 2004.

\bibitem{seideman_calculation_1992}
Tamar Seideman and William~H. Miller.
\newblock Calculation of the cumulative reaction probability via a discrete
  variable representation with absorbing boundary conditions.
\newblock {\em The Journal of Chemical Physics}, 96(6):4412--4422, March 1992.

\bibitem{baer_quantum_1997}
Roi Baer and Ronnie Kosloff.
\newblock Quantum dissipative dynamics of adsorbates near metal surfaces: {A}
  surrogate {Hamiltonian} theory applied to hydrogen on nickel.
\newblock {\em The Journal of Chemical Physics}, 106(21):8862--8875, June 1997.

\bibitem{koch_femtosecond_2003}
Christiane~P. Koch, Thorsten Kl{\"u}ner, Hans-Joachim Freund, and Ronnie
  Kosloff.
\newblock Femtosecond {Photodesorption} of {Small} {Molecules} from {Surfaces}:
  {A} {Theoretical} {Investigation} from {First} {Principles}.
\newblock {\em Physical Review Letters}, 90(11):117601, March 2003.

\bibitem{galperin_current-induced_2005}
Michael Galperin and Abraham Nitzan.
\newblock Current-{Induced} {Light} {Emission} and {Light}-{Induced} {Current}
  in {Molecular}-{Tunneling} {Junctions}.
\newblock {\em Physical Review Letters}, 95(20):206802, November 2005.

\bibitem{kleinekathofer_switching_2006}
U~Kleinekath{\"o}fer, GuangQi Li, S~Welack, and M~Schreiber.
\newblock Switching the current through model molecular wires with {Gaussian}
  laser pulses.
\newblock {\em Europhysics Letters (EPL)}, 75(1):139--145, July 2006.

\bibitem{fainberg_light-induced_2007}
B.~D. Fainberg, M.~Jouravlev, and A.~Nitzan.
\newblock Light-induced current in molecular tunneling junctions excited with
  intense shaped pulses.
\newblock {\em Physical Review B}, 76(24):245329, December 2007.

\bibitem{katz_stochastic_2008}
Gil Katz, David Gelman, Mark~A. Ratner, and Ronnie Kosloff.
\newblock Stochastic surrogate {Hamiltonian}.
\newblock {\em The Journal of Chemical Physics}, 129(3):034108, July 2008.

\bibitem{subotnik_nonequilibrium_2009}
Joseph~E. Subotnik, Thorsten Hansen, Mark~A. Ratner, and Abraham Nitzan.
\newblock Nonequilibrium steady state transport via the reduced density matrix
  operator.
\newblock {\em The Journal of Chemical Physics}, 130(14):144105, April 2009.

\bibitem{rothman_nonequilibrium_2010}
Adam~E. Rothman and David~A. Mazziotti.
\newblock Nonequilibrium, steady-state electron transport with
  {N}-representable density matrices from the anti-{Hermitian} contracted
  {Schr{\"o}dinger} equation.
\newblock {\em The Journal of Chemical Physics}, 132(10):104112, March 2010.

\bibitem{volkovich_transient_2011}
Roie Volkovich and Uri Peskin.
\newblock Transient dynamics in molecular junctions: {Coherent} bichromophoric
  molecular electron pumps.
\newblock {\em Physical Review B}, 83(3):033403, January 2011.

\bibitem{renaud_time-dependent_2011}
N.~Renaud, M.~A. Ratner, and C.~Joachim.
\newblock A {Time}-{Dependent} {Approach} to {Electronic} {Transmission} in
  {Model} {Molecular} {Junctions}, February 2011.

\bibitem{peskin_coherently_2012}
Uri Peskin and Michael Galperin.
\newblock Coherently controlled molecular junctions.
\newblock {\em The Journal of Chemical Physics}, 136(4):044107, January 2012.

\bibitem{nguyen_how_2015}
Triet~S. Nguyen, Ravindra Nanguneri, and John Parkhill.
\newblock How electronic dynamics with {Pauli} exclusion produces
  {Fermi}-{Dirac} statistics.
\newblock {\em The Journal of Chemical Physics}, 142(13):134113, April 2015.

\bibitem{baer_ab_2003}
Roi Baer and Daniel Neuhauser.
\newblock Ab initio electrical conductance of a molecular wire.
\newblock {\em International Journal of Quantum Chemistry}, 91(3):524--532,
  2003.

\bibitem{baer_ab_2004}
Roi Baer, Tamar Seideman, Shahal Ilani, and Daniel Neuhauser.
\newblock Ab initio study of the alternating current impedance of a molecular
  junction.
\newblock {\em The Journal of Chemical Physics}, 120(7):3387--3396, February
  2004.

\bibitem{bushong_approach_2005}
Neil Bushong, Na~Sai, and Massimiliano Di~Ventra.
\newblock Approach to {Steady}-{State} {Transport} in {Nanoscale} {Conductors}.
\newblock {\em Nano Letters}, 5(12):2569--2572, December 2005.

\bibitem{sanchez_molecular_2006}
Cristi{\'a}n~G. S{\'a}nchez, Maria Stamenova, Stefano Sanvito, D.~R. Bowler,
  Andrew~P. Horsfield, and Tchavdar~N. Todorov.
\newblock Molecular conduction: {Do} time-dependent simulations tell you more
  than the {Landauer} approach?
\newblock {\em The Journal of Chemical Physics}, 124(21):214708, June 2006.

\bibitem{zheng_time-dependent_2007}
Xiao Zheng, Fan Wang, Chi~Yung Yam, Yan Mo, and GuanHua Chen.
\newblock Time-dependent density-functional theory for open systems.
\newblock {\em Physical Review B}, 75(19):195127, May 2007.

\bibitem{evans_dynamic_2009}
Jeremy~S. Evans and Troy~Van Voorhis.
\newblock Dynamic {Current} {Suppression} and {Gate} {Voltage} {Response} in
  {Metal}-{Molecule}-{Metal} {Junctions}.
\newblock {\em Nano Letters}, 9(7):2671--2675, July 2009.

\bibitem{ercan_tight-binding_2010}
{\.I}lke Ercan and Neal~G. Anderson.
\newblock Tight-binding implementation of the microcanonical approach to
  transport in nanoscale conductors: {Generalization} and analysis.
\newblock {\em Journal of Applied Physics}, 107(12):124318, June 2010.

\bibitem{zheng_time-dependent_2010}
Xiao Zheng, GuanHua Chen, Yan Mo, SiuKong Koo, Heng Tian, ChiYung Yam, and
  YiJing Yan.
\newblock Time-dependent density functional theory for quantum transport.
\newblock {\em The Journal of Chemical Physics}, 133(11):114101, September
  2010.

\bibitem{xing_first-principles_2010}
Yanxia Xing, Bin Wang, and Jian Wang.
\newblock First-principles investigation of dynamical properties of molecular
  devices under a steplike pulse.
\newblock {\em Physical Review B}, 82(20):205112, November 2010.

\bibitem{ke_time-dependent_2010}
San-Huang Ke, Rui Liu, Weitao Yang, and Harold~U. Baranger.
\newblock Time-dependent transport through molecular junctions.
\newblock {\em The Journal of Chemical Physics}, 132(23):234105, June 2010.

\bibitem{wang_time-dependent_2013}
Rulin Wang, Dong Hou, and Xiao Zheng.
\newblock Time-dependent density-functional theory for real-time electronic
  dynamics on material surfaces.
\newblock {\em Physical Review B}, 88(20):205126, November 2013.

\bibitem{schaffhauser_using_2016}
Philipp Schaffhauser and Stephan K{\"u}mmel.
\newblock Using time-dependent density functional theory in real time for
  calculating electronic transport.
\newblock {\em Physical Review B}, 93(3):035115, January 2016.

\bibitem{zelovich_state_2014}
Tamar Zelovich, Leeor Kronik, and Oded Hod.
\newblock State {Representation} {Approach} for {Atomistic} {Time}-{Dependent}
  {Transport} {Calculations} in {Molecular} {Junctions}.
\newblock {\em Journal of Chemical Theory and Computation}, 10(8):2927--2941,
  August 2014.

\bibitem{zelovich_molecule-lead_2015}
Tamar Zelovich, Leeor Kronik, and Oded Hod.
\newblock Molecule-{Lead} {Coupling} at {Molecular} {Junctions}: {Relation}
  between the {Real}- and {State}-{Space} {Perspectives}.
\newblock {\em Journal of Chemical Theory and Computation}, 11(10):4861--4869,
  October 2015.

\bibitem{zelovich_driven_2016}
Tamar Zelovich, Leeor Kronik, and Oded Hod.
\newblock Driven {Liouville} von {Neumann} {Approach} for {Time}-{Dependent}
  {Electronic} {Transport} {Calculations} in a {Nonorthogonal} {Basis}-{Set}
  {Representation}.
\newblock {\em The Journal of Physical Chemistry C}, 120(28):15052--15062, July
  2016.

\bibitem{zelovich_parameter-free_2017}
Tamar Zelovich, Thorsten Hansen, Zhen-Fei Liu, Jeffrey~B. Neaton, Leeor Kronik,
  and Oded Hod.
\newblock Parameter-free driven {Liouville}-von {Neumann} approach for
  time-dependent electronic transport simulations in open quantum systems.
\newblock {\em The Journal of Chemical Physics}, 146(9):092331, March 2017.

\bibitem{hod_driven_2016}
Oded Hod, Cesar~A. Rodriguez-Rosario, Tamar Zelovich, and Thomas Frauenheim.
\newblock Driven {Liouville} von {Neumann} {Equation} in {Lindblad} {Form}.
\newblock {\em The Journal of Physical Chemistry A}, 120(19):3278--3285, May
  2016.

\bibitem{gruss_landauers_2016}
Daniel Gruss, Kirill~A. Velizhanin, and Michael Zwolak.
\newblock Landauer{\textquoteright}s formula with finite-time relaxation:
  {Kramers}{\textquoteright} crossover in electronic transport.
\newblock {\em Scientific Reports}, 6(1):24514, July 2016.

\bibitem{elenewski_communication:_2017}
Justin~E. Elenewski, Daniel Gruss, and Michael Zwolak.
\newblock Communication: {Master} equations for electron transport: {The}
  limits of the {Markovian} limit.
\newblock {\em The Journal of Chemical Physics}, 147(15):151101, October 2017.

\bibitem{verzijl_applicability_2013}
C.~J.~O. Verzijl, J.~S. Seldenthuis, and J.~M. Thijssen.
\newblock Applicability of the wide-band limit in {DFT}-based molecular
  transport calculations.
\newblock {\em The Journal of Chemical Physics}, 138(9):094102, March 2013.

\bibitem{baldea_invariance_2016}
Ioan B{\^a}ldea.
\newblock Invariance of molecular charge transport upon changes of extended
  molecule size and several related issues.
\newblock {\em Beilstein Journal of Nanotechnology}, 7:418--431, March 2016.

\bibitem{covito_transient_2018}
F.~Covito, F.~G. Eich, R.~Tuovinen, M.~A. Sentef, and A.~Rubio.
\newblock Transient {Charge} and {Energy} {Flow} in the {Wide}-{Band} {Limit}.
\newblock {\em Journal of Chemical Theory and Computation}, 14(5):2495--2504,
  May 2018.

\bibitem{nitzan_chemical_2006}
A.~Nitzan.
\newblock {\em Chemical {Dynamics} in {Condensed} {Phases}: {Relaxation},
  {Transfer} and {Reactions} in {Condensed} {Molecular} {Systems}}.
\newblock Oxford {Graduate} {Texts}. OUP Oxford, 2006.

\bibitem{koentopp_density_2008}
Max Koentopp, Connie Chang, Kieron Burke, and Roberto Car.
\newblock Density functional calculations of nanoscale conductance.
\newblock {\em Journal of Physics: Condensed Matter}, 20(8):083203, February
  2008.

\bibitem{kurth_time-dependent_2005}
S.~Kurth, G.~Stefanucci, C.-O. Almbladh, A.~Rubio, and E.~K.~U. Gross.
\newblock Time-dependent quantum transport: {A} practical scheme using density
  functional theory.
\newblock {\em Physical Review B}, 72(3):035308, July 2005.

\bibitem{ochoa_energy_2016}
Maicol~A. Ochoa, Anton Bruch, and Abraham Nitzan.
\newblock Energy distribution and local fluctuations in strongly coupled open
  quantum systems: {The} extended resonant level model.
\newblock {\em Physical Review B}, 94(3):035420, July 2016.

\bibitem{morzan_electron_2017}
Uriel~N. Morzan, Francisco~F. Ram{\'i}rez, Mariano~C. Gonz{\'a}lez~Lebrero, and
  Dami{\'a}n~A. Scherlis.
\newblock Electron transport in real time from first-principles.
\newblock {\em The Journal of Chemical Physics}, 146(4):044110, January 2017.

\bibitem{bixon_intramolecular_1968}
Mordechai Bixon and Joshua Jortner.
\newblock Intramolecular {Radiationless} {Transitions}.
\newblock {\em The Journal of Chemical Physics}, 48(2):715--726, January 1968.

\bibitem{carmeli_random_1982}
Benny Carmeli, Roberto Tulman, Abraham Nitzan, and M.~H. Kalos.
\newblock Random coupling models.{IV}. {Numerical} investigation of the
  dependence on the random coupling distribution and on the initial phases.
\newblock {\em Chemical Physics}, 72(3):363 -- 369, 1982.

\end{thebibliography}

\appendix

\section{Energy Resolved Currents Calculation\label{sec:Energy-Resolved-Currents}}

Energy resolved currents can be evaluated via the time derivatives
of the various lead levels populations. Here, we show how the corresponding
expression (Eq. \ref{eq: current at dot/lead interface}) is obtained
for the resonant level system discussed in the main text. For a system
composed of a single dot level and a manifold of lead states that
are directly coupled to an implicit bath (Eq. \ref{eq: H}), the DLvN
equation of motion is given by the following matrix representation
written in the basis of eigenstates of the isolated dot and lead sections
(Eq. \ref{eq: DLvN}):

\begin{equation}
\frac{\text{d}}{\text{d}t}\left(\begin{array}{cc}
\sigma_{d}\left(t\right) & \hat{\sigma}_{dL}\left(t\right)\\
\hat{\sigma}_{Ld}\left(t\right) & \hat{\sigma}_{L}\left(t\right)
\end{array}\right)=-\frac{i}{\hbar}\left[\left(\begin{array}{cc}
\varepsilon_{d}\left(t\right) & \hat{V}_{dL}\\
\hat{V}_{Ld} & \hat{H}_{L}
\end{array}\right),\left(\begin{array}{cc}
\sigma_{d}\left(t\right) & \hat{\sigma}_{dL}\left(t\right)\\
\hat{\sigma}_{Ld}\left(t\right) & \hat{\sigma}_{L}\left(t\right)
\end{array}\right)\right]-\Gamma\left(\begin{array}{cc}
0 & \frac{1}{2}\hat{\sigma}_{dL}(t)\\
\frac{1}{2}\hat{\sigma}_{Ld}(t) & \hat{\sigma}_{L}(t)-\hat{\sigma}_{L}^{0}
\end{array}\right).\label{eq: DLvN appendix currents}
\end{equation}

Evaluating the commutator on the right hand side of Eq. \ref{eq: DLvN appendix currents},
while taking into consideration that there is a single dot level,
gives:

{\small{}
\begin{align}
 & \frac{\text{d}}{\text{d}t}\left(\begin{array}{cc}
\sigma_{d}\left(t\right) & \hat{\sigma}_{dL}\left(t\right)\\
\hat{\sigma}_{Ld}\left(t\right) & \hat{\sigma}_{L}\left(t\right)
\end{array}\right)=-\Gamma\left(\begin{array}{cc}
0 & \frac{1}{2}\hat{\sigma}_{dL}(t)\\
\frac{1}{2}\hat{\sigma}_{Ld}(t) & \hat{\sigma}_{L}(t)-\hat{\sigma}_{L}^{0}
\end{array}\right)\nonumber \\
 & -\frac{i}{\hbar}\left(\begin{array}{cc}
\hat{V}_{dL}\hat{\sigma}_{Ld}\left(t\right)-\hat{\sigma}_{dL}\left(t\right)\hat{V}_{Ld} & \varepsilon_{d}\left(t\right)\hat{\sigma}_{dL}\left(t\right)+\hat{V}_{dL}\hat{\sigma}_{L}\left(t\right)-\sigma_{d}\left(t\right)\hat{V}_{dL}-\hat{\sigma}_{dL}\left(t\right)\hat{H}_{L}\\
\hat{V}_{Ld}\sigma_{d}\left(t\right)+\hat{H}_{L}\hat{\sigma}_{Ld}\left(t\right)-\hat{\sigma}_{Ld}\left(t\right)\varepsilon_{d}\left(t\right)-\hat{\sigma}_{L}\left(t\right)\hat{V}_{Ld} & \hat{V}_{Ld}\hat{\sigma}_{dL}\left(t\right)+\hat{H}_{L}\hat{\sigma}_{L}\left(t\right)-\hat{\sigma}_{Ld}\left(t\right)\hat{V}_{dL}-\hat{\sigma}_{L}\left(t\right)\hat{H}_{L}
\end{array}\right).
\end{align}
}{\small\par}

Hence, the dynamics of the lead section is given by:

\begin{equation}
\frac{\text{d}}{\text{d}t}\hat{\sigma}_{L}\left(t\right)=-\frac{i}{\hbar}\left[\hat{H}_{L},\hat{\sigma}_{L}\left(t\right)\right]-\frac{i}{\hbar}\left[\hat{V}_{Ld}\hat{\sigma}_{dL}\left(t\right)-\hat{\sigma}_{Ld}\left(t\right)\hat{V}_{dL}\right]-\Gamma\left(\hat{\sigma}_{L}(t)-\hat{\sigma}_{L}^{0}\right).
\end{equation}

From this we can calculate the rate of population variation in a given
lead level $l$ as:

\begin{equation}
\frac{\text{d}}{\text{d}t}\sigma_{ll}\left(t\right)=-\frac{i}{\hbar}\sum_{l^{'}=0}^{N_{l}}\left[H_{ll^{'}}\sigma_{l^{'}l}\left(t\right)-\sigma_{ll^{'}}\left(t\right)H_{l^{'}l}\right]-\frac{i}{\hbar}\left[V_{ld}\sigma_{dl}\left(t\right)-\sigma_{ld}\left(t\right)V_{dl}\right]-\Gamma\left(\sigma_{ll}(t)-\sigma_{ll}^{0}\right).\label{eq: d/dt sigmal_ll app currents}
\end{equation}

In the representation of the eigenbasis of the isolated dot and lead
states, $\hat{H}_{L}$ is diagonal and thus the first term on the
right hand side of Eq. \ref{eq: d/dt sigmal_ll app currents} vanishes.
The remaining two terms can be identified as the particle current
flowing between lead state $l$ and the dot or the implicit bath,
respectively. Focusing on the second term and taking into account
the fact that $\hat{H}$ and $\hat{\sigma}$ are Hermitian matrices,
such that $V_{dl}=V_{ld}^{*}$ and $\sigma_{ld}\left(t\right)=\sigma_{dl}^{*}\left(t\right)$,
we arrive at the expression for the current flowing between from dot
to lead level $l$ (Eq. \ref{eq: current at dot/lead interface} in
the main text):

\begin{equation}
J_{dL}\left(\varepsilon_{l},t\right)=-\frac{i}{\hbar}\left[V_{ld}\sigma_{dl}\left(t\right)-\sigma_{dl}^{*}\left(t\right)V_{ld}^{*}\right]=\frac{2}{\hbar}\Im\left(V_{ld}\sigma_{dl}\left(t\right)\right).\label{eq: current d/L appendix currents}
\end{equation}

If we neglect the lead level width due to its coupling to the bath
and the dot, $J_{dL}\left(\varepsilon_{l},t\right)$ represents the
energy resolved current flowing from the dot to the lead at energy
$\varepsilon_{l}$. Correspondingly, the particle current flowing
at energy $\varepsilon_{l}$ from the lead to the implicit bath is
given by (Eq. \ref{eq: current at lead/bath} in the main text):

\begin{equation}
J_{LB}\left(\varepsilon_{l},t\right)=\Gamma\left(\sigma_{ll}(t)-\sigma_{ll}^{0}\right).
\end{equation}

The same holds true when the model system is decomposed into the dot,
lead, and driven lead sections (Eq. \ref{eq: H for dLK} in the main
text), where Eq. \ref{eq: current d/L appendix currents} remains
valid for the current flowing between the dot and lead level $l$,
and a similar expression is obtained for the current flowing from
the lead into driven lead level $k$:

\begin{equation}
J_{LK}\left(\varepsilon_{k},t\right)=\frac{2}{\hbar}\sum_{l=0}^{N_{l}}\Im\left(V_{kl}\sigma_{lk}\left(t\right)\right).
\end{equation}

\section{Sylvester Equation For the Equilibrium Density Matrix\label{sec:Sylvester-equation-for}}

Within the DLvN approach, the equilibrium state of a single lead setup
and the steady-state of a multi-lead setup can be obtained by setting
$\frac{d\hat{\sigma}\left(t\right)}{dt}=0$. As mentioned in the main
text, for the former the obtained equilibrium is appoximate as the
equation of motion drives only the lead section, rather than the entire
system (dot+lead) towards equilibrium. In principle, equilibrium can
be reached by running the dynamics until all transient effects relax.
This, however, may prove to be computationally quite inefficient,
especially when the initial conditions are far from equilibrium. An
alternative can be to formulate an equation that directly solves for
$\frac{d\hat{\sigma}\left(t\right)}{dt}=0$ \cite{subotnik_nonequilibrium_2009}.
To this end, in the case of a single lead setup, we define projection
operators on the dot and on the lead section as

\begin{equation}
\hat{P}=\left(\begin{array}{cc}
1 & \hat{0}\\
\hat{0} & \hat{0}
\end{array}\right),\ \hat{Q}=\left(\begin{array}{cc}
0 & \hat{0}\\
\hat{0} & \hat{1}
\end{array}\right),\label{eq: P projection for lead-dot}
\end{equation}

corresponding to the matrix representation in the basis of the isolated
dot and lead eigenfunctions. With these, the driving term in Eq. \ref{eq: DLvN}
of the main text can be decomposed as follows:

\begin{align}
-\Gamma\cdot\left(\begin{array}{cc}
0 & \frac{1}{2}\hat{\sigma}_{d,L}\left(t\right)\\
\frac{1}{2}\hat{\sigma}_{L,d}\left(t\right) & \hat{\sigma}_{L}\left(t\right)-\hat{\sigma}_{L}^{0}
\end{array}\right) & =-\frac{1}{2}\Gamma\cdot\underbrace{\left(\begin{array}{cc}
0 & \hat{\sigma}_{d,L}\left(t\right)\\
\hat{0} & \hat{0}
\end{array}\right)}_{\hat{P}\hat{\sigma}\left(t\right)\hat{Q}}-\frac{1}{2}\Gamma\cdot\underbrace{\left(\begin{array}{cc}
0 & \hat{0}\\
\hat{\sigma}_{L,d}\left(t\right) & \hat{0}
\end{array}\right)}_{\hat{Q}\hat{\sigma}\left(t\right)\hat{P}}\nonumber \\
 & -\Gamma\cdot\underbrace{\left(\begin{array}{cc}
0 & \hat{0}\\
\hat{0} & \hat{\sigma}_{L}\left(t\right)
\end{array}\right)}_{\hat{Q}\hat{\sigma}\left(t\right)\hat{Q}}+\Gamma\cdot\underbrace{\left(\begin{array}{cc}
0 & \hat{0}\\
\hat{0} & \hat{\sigma}_{L}^{0}
\end{array}\right)}_{\hat{Q}\hat{\sigma}^{0}\hat{Q}},\label{eq: gm term opening}
\end{align}

where in the last term

\begin{equation}
\hat{\sigma}^{0}=\left(\begin{array}{cc}
\sigma_{d}^{0} & \hat{0}\\
\hat{0} & \hat{\sigma}_{L}^{0}
\end{array}\right),\label{eq: target density matrix}
\end{equation}

and the target equilibrium occupation of the dot, $\sigma_{d}^{0}$,
does not (and should not) appear in the final expression for the driving
term. Nullifying the left hand side of the DLvN equation of motion
(Eq. \ref{eq: DLvN}) with a time-independent Hamiltonian thus gives
the following equation for the equilibrium density matrix of the system,
$\hat{\sigma}^{eq}$:

\begin{equation}
\frac{\text{d}}{\text{d}t}\hat{\sigma}^{eq}=-\frac{i}{\hbar}\hat{H}\hat{\sigma}^{eq}+\frac{i}{\hbar}\hat{\sigma}^{eq}\hat{H}-\frac{1}{2}\Gamma\hat{P}\hat{\sigma}^{eq}\hat{Q}-\frac{1}{2}\Gamma\hat{Q}\hat{\sigma}^{eq}\hat{P}-\Gamma\hat{Q}\hat{\sigma}^{eq}\hat{Q}+\Gamma\hat{Q}\hat{\sigma}^{0}\hat{Q}=\hat{0},\label{eq: DLvN zero for lead-dot}
\end{equation}

which can be rearranged as:

\begin{equation}
-\frac{i}{\hbar}\hat{H}\hat{\sigma}^{eq}+\frac{i}{\hbar}\hat{\sigma}^{eq}\hat{H}-\frac{1}{2}\Gamma\left(\hat{P}+\hat{Q}\right)\hat{\sigma}^{eq}\hat{Q}-\frac{1}{2}\Gamma\hat{Q}\hat{\sigma}^{eq}\left(\hat{P}+\hat{Q}\right)+\Gamma\hat{Q}\hat{\sigma}^{0}\hat{Q}=\hat{0}.
\end{equation}

Since the sum of $\hat{P}$ and $\hat{Q}$ gives the unity matrix,
$\hat{1}$, this results in:

\begin{equation}
\left(\frac{i}{\hbar}\hat{H}+\frac{1}{2}\Gamma\hat{Q}\right)\hat{\sigma}^{eq}-\hat{\sigma}^{eq}\left(\frac{i}{\hbar}\hat{H}-\frac{1}{2}\Gamma\hat{Q}\right)=\Gamma\hat{Q}\hat{\sigma}^{0}\hat{Q}.
\end{equation}

Defining

\begin{equation}
\hat{A}=\frac{i}{\hbar}\hat{H}+\frac{1}{2}\Gamma\hat{Q}=\frac{i}{\hbar}\left(\begin{array}{cc}
\varepsilon_{d} & \hat{V}_{dL}\\
\hat{V}_{Ld} & \hat{H}_{L}-\frac{i\hbar}{2}\Gamma\hat{1}
\end{array}\right),
\end{equation}

\begin{equation}
\hat{B}=\frac{i}{\hbar}\hat{H}-\frac{1}{2}\Gamma\hat{Q}=\frac{i}{\hbar}\left(\begin{array}{cc}
\varepsilon_{d} & \hat{V}_{dL}\\
\hat{V}_{Ld} & \hat{H}_{L}+\frac{i\hbar}{2}\Gamma\hat{1}
\end{array}\right),
\end{equation}

and

\begin{equation}
\hat{C}=\Gamma\hat{Q}\hat{\sigma}^{0}\hat{Q}=\left(\begin{array}{cc}
0 & \hat{0}\\
\hat{0} & \Gamma\hat{\sigma}_{L}^{0}
\end{array}\right),
\end{equation}

the following Sylvester equation for the equilibrium density matrix
is obtained \cite{subotnik_nonequilibrium_2009}:

\begin{equation}
\hat{A}\hat{\sigma}^{eq}-\hat{\sigma}^{eq}\hat{B}=\hat{C}.\label{eq: in sylvester form}
\end{equation}

A similar Sylvester equation can be derived when the system is decomposed
into the dot, lead, and driven lead sections. In the basis of eigenfunctions
of these isolated sections the corresponding projection operators
are written as:

\begin{equation}
\hat{P}=\left(\begin{array}{ccc}
1 & \hat{0} & \hat{0}\\
\hat{0} & \hat{0} & \hat{0}\\
\hat{0} & \hat{0} & \hat{0}
\end{array}\right),\ \hat{Q}=\left(\begin{array}{ccc}
0 & \hat{0} & \hat{0}\\
\hat{0} & \hat{1} & \hat{0}\\
\hat{0} & \hat{0} & \hat{0}
\end{array}\right),\ \hat{R}=\left(\begin{array}{ccc}
0 & \hat{0} & \hat{0}\\
\hat{0} & \hat{0} & \hat{0}\\
\hat{0} & \hat{0} & \hat{1}
\end{array}\right).\label{eq: P projection-1}
\end{equation}

The driving term (see Eq. \ref{eq: DLvN for dLK}) can be decomposed
in terms of these projection operators as follows:

\begin{align}
 & -\Gamma\cdot\left(\begin{array}{ccc}
0 & \hat{0} & \frac{1}{2}\hat{\sigma}_{dK}\left(t\right)\\
\hat{0} & \hat{0} & \frac{1}{2}\hat{\sigma}_{LK}\left(t\right)\\
\frac{1}{2}\hat{\sigma}_{Kd}\left(t\right) & \frac{1}{2}\hat{\sigma}_{KL}\left(t\right) & \hat{\sigma}_{K}\left(t\right)-\hat{\sigma}_{K}^{0}
\end{array}\right)\nonumber \\
 & =-\frac{1}{2}\Gamma\cdot\underbrace{\left(\begin{array}{ccc}
0 & \hat{0} & \hat{0}\\
\hat{0} & \hat{0} & \hat{0}\\
\hat{\sigma}_{Kd}\left(t\right) & \hat{0} & \hat{0}
\end{array}\right)}_{\hat{R}\hat{\sigma}\left(t\right)\hat{P}}-\frac{1}{2}\Gamma\cdot\underbrace{\left(\begin{array}{ccc}
0 & \hat{0} & \hat{0}\\
\hat{0} & \hat{0} & \hat{0}\\
\hat{0} & \hat{\sigma}_{KL}\left(t\right) & \hat{0}
\end{array}\right)}_{\hat{R}\hat{\sigma}\left(t\right)\hat{Q}}\nonumber \\
 & -\frac{1}{2}\Gamma\cdot\underbrace{\left(\begin{array}{ccc}
0 & \hat{0} & \hat{\sigma}_{dK}\left(t\right)\\
\hat{0} & \hat{0} & \hat{0}\\
\hat{0} & \hat{0} & \hat{0}
\end{array}\right)}_{\hat{P}\hat{\sigma}\left(t\right)\hat{R}}-\frac{1}{2}\Gamma\cdot\underbrace{\left(\begin{array}{ccc}
0 & \hat{0} & \hat{0}\\
\hat{0} & \hat{0} & \hat{\sigma}_{LK}\left(t\right)\\
\hat{0} & \hat{0} & \hat{0}
\end{array}\right)}_{\hat{Q}\hat{\sigma}\left(t\right)\hat{R}}-\Gamma\cdot\underbrace{\left(\begin{array}{ccc}
0 & \hat{0} & \hat{0}\\
\hat{0} & \hat{0} & \hat{0}\\
\hat{0} & \hat{0} & \hat{\sigma}_{K}\left(t\right)
\end{array}\right)}_{\hat{R}\hat{\sigma}\left(t\right)\hat{R}}+\Gamma\cdot\underbrace{\left(\begin{array}{ccc}
0 & \hat{0} & \hat{0}\\
\hat{0} & \hat{0} & \hat{0}\\
\hat{0} & \hat{0} & \hat{\sigma}_{K}^{0}
\end{array}\right)}_{\hat{R}\hat{\sigma}^{0}\hat{R}},
\end{align}

where, in the last term,

\begin{equation}
\hat{\sigma}^{0}=\underbrace{\left(\begin{array}{ccc}
\sigma_{d}^{0} & \hat{0} & \hat{0}\\
\hat{0} & \hat{\sigma}_{L}^{0} & \hat{0}\\
\hat{0} & \hat{0} & \hat{\sigma}_{K}^{0}
\end{array}\right)},\label{eq: DLvN zero for lead-dot-1-1-1}
\end{equation}

and the target equilibrium occupation of the dot $\left(\sigma_{d}^{0}\right)$
and the lead $\left(\sigma_{L}^{0}\right)$ do not (and should not)
appear in the final expression for the driving term. Nullifying the
left hand side of the DLvN equation of motion (Eq. \ref{eq: DLvN for dLK})
with a time-independent Hamiltonian thus gives the following equation
for the equilibrium density matrix of the system, $\hat{\sigma}^{eq}$:

\begin{align}
\frac{\text{d}}{\text{d}t}\hat{\sigma}^{eq}=-\frac{i}{\hbar}\tilde{\hat{H}}\hat{\sigma}^{eq}+\frac{i}{\hbar}\hat{\sigma}^{eq}\tilde{\hat{H}}-\frac{1}{2}\Gamma\hat{R}\hat{\sigma}^{eq}\hat{P}-\frac{1}{2}\Gamma\hat{R}\hat{\sigma}^{eq}\hat{Q}-\frac{1}{2}\Gamma\hat{P}\hat{\sigma}^{eq}\hat{R}-\frac{1}{2}\Gamma\hat{Q}\hat{\sigma}^{eq}\hat{R}-\Gamma\hat{R}\hat{\sigma}^{eq}\hat{R}+\Gamma\hat{R}\hat{\sigma}^{0}\hat{R}=\hat{0}
\end{align}

which can be rearranged as:

\begin{equation}
-\frac{i}{\hbar}\tilde{\hat{H}}\hat{\sigma}^{eq}+\frac{i}{\hbar}\hat{\sigma}^{eq}\tilde{\hat{H}}-\frac{1}{2}\Gamma\hat{R}\hat{\sigma}^{eq}\left(\hat{P}+\hat{Q}+\hat{R}\right)-\frac{1}{2}\Gamma\left(\hat{P}+\hat{Q}+\hat{R}\right)\hat{\sigma}^{eq}\hat{R}+\Gamma\hat{R}\hat{\sigma}^{0}\hat{R}=\hat{0}.\label{eq: DLvN zero for lead-dot-2-2}
\end{equation}

Since the sum of $\hat{P}+\hat{Q}+\hat{R}=\hat{1}$ this results in:

\begin{equation}
\left(\frac{i}{\hbar}\tilde{\hat{H}}+\frac{1}{2}\Gamma\hat{R}\right)\hat{\sigma}^{eq}-\hat{\sigma}^{eq}\left(\frac{i}{\hbar}\tilde{\hat{H}}-\frac{1}{2}\Gamma\hat{R}\right)=\Gamma\hat{R}\hat{\sigma}^{0}\hat{R}.\label{eq: DLvN zero for lead-dot-2-1-5}
\end{equation}

Defining

\begin{equation}
\hat{A}=\frac{i}{\hbar}\tilde{\hat{H}}+\frac{1}{2}\Gamma\hat{R}=\frac{i}{\hbar}\left(\begin{array}{ccc}
\varepsilon_{d} & \tilde{\hat{V}}_{dL} & \hat{0}\\
\tilde{\hat{V}}_{Ld} & \tilde{\hat{H}}_{L} & \tilde{\hat{V}}_{LK}\\
\hat{0} & \tilde{\hat{V}}_{KL} & \tilde{\hat{H}}_{K}-\frac{i\hbar}{2}\Gamma\hat{1}
\end{array}\right),\label{eq: DLvN zero for lead-dot-2-1-1-1}
\end{equation}

\begin{equation}
\hat{B}=\frac{i}{\hbar}\tilde{\hat{H}}-\frac{1}{2}\Gamma\hat{R}=\frac{i}{\hbar}\left(\begin{array}{ccc}
\varepsilon_{d} & \tilde{\hat{V}}_{dL} & \hat{0}\\
\tilde{\hat{V}}_{Ld} & \tilde{\hat{H}}_{L} & \tilde{\hat{V}}_{LK}\\
\hat{0} & \tilde{\hat{V}}_{KL} & \tilde{\hat{H}}_{K}+\frac{i\hbar}{2}\Gamma\hat{1}
\end{array}\right),\label{eq: DLvN zero for lead-dot-2-1-2-1}
\end{equation}

and

\begin{equation}
\hat{C}=\Gamma\hat{R}\hat{\sigma}^{0}\hat{R}=\left(\begin{array}{ccc}
0 & \hat{0} & \hat{0}\\
\hat{0} & \hat{0} & \hat{0}\\
\hat{0} & \hat{0} & \Gamma\hat{\sigma}_{K}^{0}
\end{array}\right),\label{eq: DLvN zero for lead-dot-2-1-3-1}
\end{equation}

we arrive at a Sylvester equation of the same structure as Eq. \ref{eq: in sylvester form}
above.
\end{document}